\documentclass[preprint,showpacs,preprintnumbers,amsmath,amssymb]{revtex4}
\usepackage{graphicx}
\usepackage{dcolumn}
\usepackage{bm}

\newcommand{\be}{\begin{equation}}
\newcommand{\en}{\end{equation}}
\newcommand{\bea}{\begin{eqnarray}}
\newcommand{\ena}{\end{eqnarray}}

\begin{document}

\title{General dissipative coefficient in warm intermediate
inflation in loop quantum cosmology in light of Planck and BICEP2}
\author{Ram\'on Herrera}

\email{ramon.herrera@ucv.cl} \affiliation{ Instituto de
F\'{\i}sica, Pontificia Universidad Cat\'{o}lica de Valpara\'{\i}so,
Casilla 4059, Valpara\'{\i}so, Chile.}

\author{Marco Olivares}
\email{marco.olivares@ucv.cl} \affiliation{ Instituto de F\'{\i}sica,
Pontificia Universidad Cat\'{o}lica de Valpara\'{\i}so, Casilla
4059, Valpara\'{\i}so, Chile.}

\author{Nelson Videla}
\email{nelson.videla@ucv.cl} \affiliation{ Instituto de
F\'{\i}sica, Pontificia Universidad Cat\'{o}lica de
Valpara\'{\i}so, Casilla 4059, Valpara\'{\i}so, Chile.}

\date{\today}

\begin{abstract}
 In this paper, we study a warm intermediate inflationary model
with a general form for the dissipative coefficient
$\Gamma(T,\phi)=C_\phi\,T^{m}/\phi^{m-1}$  in the context of loop
quantum cosmology. We examine this model in the weak and strong
dissipative regimes. In general, we discuss  in great detail the
characteristics of this model in the slow-roll approximation.
Also, we assume that the modifications to  perturbation equations
result  exclusively from Hubble rate. In this approach, we use
recent astronomical observations from Planck and BICEP2
experiments to restrict the parameters in our model.
\end{abstract}

\maketitle

\section{Introduction}

In  cosmology our concepts  concerning the early universe have
introduced a new ingredient, the inflationary phase of the
universe, which provides an attractive approach for resolving some
of the problems of the standard  model of the universe, as the
flatness, horizon, etc. \cite{R1,R102,R103,R104,R105,R106}. Also,
it is well known that inflation provides a graceful mechanism to
clarify the large-scale structure \cite{R2,R202,R203,R204,R205}
and the observed anisotropy of the cosmic microwave background
(CMB) radiation \cite{astro,astro2,astro202,Planck}. Recently, the
effects from BICEP2 experiment of gravitational waves in the
B-mode has been analyzed in Ref. \cite{Ade:2014xna}. An important
observational quantity obtained in this  experiment, is the
tensor-to-scalar ratio $r$, which $r=0.2^{+0.07}_{-0.05}$ (68 $\%$
C.L.) and takes out the value $r=0$ ( at a significance of 7.0
$\sigma$). Therefore, the tensor mode should not be neglected.

On the other hand,  warm inflation  differs from the cold
inflation since  evades  the reheating period at the end of the
accelerated evolution of the universe \cite{warm}. During warm
inflation the process of radiation production could take place
under strong enough dissipation
\cite{warm,taylorberera,taylorberera02,taylorberera03,taylorberera04,taylorberera05,
taylorberera06,taylorberera07,taylorberera08,taylorberera09}. In
this form, the dissipative effects are important and these emerge
from a friction term since the inflaton field is dissipated into a
thermal bath. Also, an  interesting feature of the warm
inflationary model is that the thermal fluctuations constitute a
dominant character  in producing the primary density fluctuations
essential for Large-Scale Structure (LSS)
formation \cite{62526,6252602,6252603,6252604,1126}.

In the context of the  dissipative effects,  a fundamental
quantity is the dissipation coefficient $\Gamma$. In particular,
for the scenario of low-temperature, the parameter $\Gamma$ was
analyzed in supersymmetric models. In these models, there is  a
scalar field together with multiplets of heavy and light fields
that give different expressions for the dissipation coefficient,
see
Refs.\cite{26,28,2802,Zhang:2009ge,BasteroGil:2011xd,BasteroGil:2012cm}.
A general form for the dissipative coefficient $\Gamma$, is given
by \cite{Zhang:2009ge,BasteroGil:2011xd}.

\begin{equation}
\Gamma=C_{\phi}\,\frac{T^{m}}{\phi^{m-1}}, \label{G}%
\end{equation}
where  the constant  $C_\phi$ is related with  the dissipative
microscopic dynamics and the constant $m$ is an integer. Various
elections of $\Gamma$ or equivalently of $m$ have been assumed in
the written works \cite{Zhang:2009ge,BasteroGil:2011xd}. In
special, for the value of $m=3$,  $C_\phi$ corresponds  to
   $C_{\phi}=0.64\,h^{4}\,\mathcal{N}$ in which ${\mathcal{N}}%
={\mathcal{N}}_{\chi}{\mathcal{N}}_{decay}^{2}$. Here,
$\mathcal{N}_{\chi}$ is the multiplicity of the $X$ superfield and
${\mathcal{N}}_{decay}$ represents   the number of decay channels
available in $X$'s
decay \cite{26,27,Berera:2008ar,BasteroGil:2010pb}. For the
special case $m=1$, i.e., the dissipation coefficient
$\Gamma\propto\,T$  corresponds to the high temperature
supersymmetry (SUSY) case. For the value  $m=0$, then
$\Gamma\propto\phi$ and the dissipation coefficient  represents
an exponentially decaying propagator in the high temperature SUSY
model. For the case $m=-1$, i.e., $\Gamma\propto\phi^2/T$, it
agrees with  the non-SUSY case \cite{28,PRD}.


On the other hand, Loop Quantum Gravity (LQG) is a proceeding  of
nonperturbative background autonomous approach to quantize gravity
\cite{5}. In  LQC the geometry  is discrete and the continuum
space-time is found from quantum geometry in a large eigenvalue
limit (see, Refs. \cite{Ashtekar:2011ni,6,7, 8, 9}). Different cosmological models
have been studied, in particular the Friedmann-Robertson-Walker
(FRW) model \cite{AA}. Here, the loop quantum effect modifies the
Friedmann equation  by adding  a correction term in the energy
density, specifically $\rho^2$ at the scale when $\rho$ becomes
similar to a critical density $\rho_c\approx 0.82\,G^{-2}$ ($G$ is
the Newton's gravitational constant)\cite{SinghMFE, As}. In this way, the
effective Friedmann equation becomes \be
 \label{newfried}
 H^2=\frac{\kappa}{3}\,\rho\,\left[1-\frac{\rho}{\rho_{c}}\right],
 \en
where  $H=\dot{a}/a$ is the Hubble parameter, $a$ is the scale
factor, $\kappa=8\pi G$, $\rho$ is the total energy density,
$\rho_{c}=\sqrt{3}\,\rho_{p}/(16\pi^2\gamma^3)$ is the critical
loop quantum density, and $\rho_{p}=G^{-2}$ is the Planck density.
We note that a rigorous numerical test of the
Eq.(\ref{newfried}) have been performed recently in
Ref.\cite{numMFE}.

The inflationary universe model in the context  of LQC has been
analyzed in Refs. \cite{Singhinf, good}.  In particular, the inflationary model has
 been studied in great detail for power-law and multiple fields in the context of LQC \cite{Ranken:2012hp},
 while in the Ref.\cite{Gupt:2013swa} the authors have studied different isotropic and  anisotropic  space-times for avoiding singularities in LQC. By the other hand,
 the model of  the warm inflation in LQC scenario was studied in
Ref. \cite{Herrera:2010yg}, in which the author studied the
inflationary scenario described by a standard scalar field coupled
to radiation, see also Ref. \cite{nn}. For a review of
inflationary LQC models, see Refs. \cite{agre1,
agre2,Xiao:2011mv,int1,int2}

On the other hand,  exact solutions in inflationary models can be
obtained from an  exponential potential,  frequently  called
power-law inflation. Here,  the scale factor has an expansion
power law type, where $a(t)\sim t^{p}$, where the constant
$p>1$ \cite{power}. As well, an exact solution can be found by
using a constant scalar potential which is often called de Sitter
inflationary universe \cite{R1}. Nevertheless, exact solutions can
also be found  from intermediate inflation \cite{Barrow1}. In this
inflationary model, the scale factor growths as
\begin{equation}
a(t)=\exp[\,A\,t^{f}],  \label{at}
\end{equation}
where $A$ and $f$ are two constants; $A>0$ and the value of $f$
varies between $0<f<1$ \cite{Barrow1}.  In intermediate inflation
the evolution of the scale factor, $a(t)$, is slower than
de-Sitter expansion, but quicker  than power law, hence the name
``intermediate". This intermediate evolution  was originally
elaborated  as an exact solution, but this model may be best
explained from the slow-roll approximation. In the slow-roll
approximation, it is possible to obtain a spectral index $n_s\sim
1$ and for the special value of $f=2/3$, the spectral index
correspond to Harrizon-Zel'dovich spectrum, where $n_s=1$. Also,
the quantity obtained in this   model, for the tensor-to-scalar
ratio is $r\neq 0$ \cite{ratior,Barrow3}.



Thus the goal of the paper is to study an evolving intermediate
scale factor during  warm  inflation in the framework of LQC model
together with a generalized form of dissipative coefficient
$\Gamma$. We will study the warm intermediate inflationary model
in LQC for  different values of $m$, and also we will consider
this model for two regimes, the weak and the strong dissipative
scenarios. In the context of the cosmological perturbations,
we will consider for simplicity the procedure of
Refs.\cite{good,Herrera:2010yg,g1,nn} for warm inflation in LQC,
where the perturbation equations arise only from Hubble rate.
 Also, we only study the standard inflation scenario,
that occurs after the superinflation epoch. For a review of
superinflation epoch in LQC, see Refs. \cite{good,g1}.

The outline of the paper is the follows: The next section presents
the basic equations for warm inflation in the framework of LQC
model. In the sections III and IV, we discuss the weak and strong
dissipative regimes in the intermediate  model. In both sections,
we give explicit expressions for the scalar field, the dissipative
coefficient, the scalar potential, the scalar power spectrum and
the tensor-to-scalar ratio. Also, the Planck and BICEP2 data are used to
constrain the parameters in both regimes. Finally, our conclusions
are presented in section V. We use units in which $c=\hbar=1$.

\section{Warm-LQC Inflation: Basic equations.\label{secti}}

We consider that during warm inflation, the universe is filled
with a self-interacting scalar field of energy density
$\rho_{\phi}$ and a radiation field with energy density
$\rho_{\gamma}$.  In fact, the total energy density of the
universe $\rho$ is given by $\rho=\rho_{\phi}+\rho_{\gamma}$.

In the following, we will regard  that the energy density
associated  to the standard scalar field $\rho_{\phi}$ is given by
$\rho_{\phi}=\dot{\phi}^{2}/2+V(\phi)$ and the pressure as
$P_{\phi}=\dot{\phi}^{2}/2-V(\phi)$. Here, $V(\phi)$ represents
the effective  potential. Dots mean derivatives with respect to
time.

The evolution equations for $\rho_{\phi}$ and $\rho_{\gamma}$ in warm
inflation are given by \cite{warm}
\begin{equation}
\dot{\rho_{\phi}}+3\,H\,(\rho_{\phi}+P_{\phi})=-\Gamma\;\;\dot{\phi}^{2},
\label{key_01}%
\end{equation}
and
\begin{equation}
\dot{\rho}_{\gamma}+4H\rho_{\gamma}=\Gamma\dot{\phi}^{2}, \label{key_02}%
\end{equation}
where, we recall that $\Gamma$, where $\Gamma>0$, is the dissipation
coefficient and it is responsible of the decay of the
field $\phi$ into radiation. This dissipation coefficient  can be
established to be a constant or a function of the temperature of
the thermal bath $\Gamma(T)$, the scalar field $\Gamma(\phi)$, or
both $\Gamma(T,\phi)$ \cite{warm}.

During  the evolution of warm inflation,  the energy density
related to the  field $\phi$ dominates over the energy density
$\rho_\gamma$ \cite{warm,62526,6252602,6252603,6252604} and, then
the Eq.(\ref{newfried}) results
\begin{equation}
H^{2}\approx\frac{\kappa}{3}\,\rho_{\phi}\,\left[1-\frac{\rho_\phi}{\rho_c}\right]=\frac{\kappa}{3}\,
\left(\frac{\dot{\phi}}{2}+V(\phi)\right)\,\left[1-\frac{\frac{\dot{\phi}}{2}+V(\phi)}{\rho_c}\right]. \label{inf2}%
\end{equation}

Considering Eqs.(\ref{key_01}) and (\ref{inf2}), we find
\begin{equation}
\dot{\phi}^{2}= \frac{2(-\dot{H})}{\kappa(1+R)} \left[  1-{\frac{12 H^{2}%
}{\kappa\rho_{c}}}\right]  ^{-1/2}, \label{inf3}%
\end{equation}
where the ratio between $\Gamma$ and the Hubble parameter $H$ is
denoted by  $R=\frac{\Gamma}{3H}$. In this sense,
 for the case of the weak or strong
dissipation regime, we make $R<1$ or $R>1$, respectively.

We consider that during warm inflation the radiation production is
quasi-stable, in which $\dot{\rho}_{\gamma}\ll4H\rho_{\gamma}$ and
$\dot{\rho }_{\gamma}\ll\Gamma\dot{\phi}^{2}$, see
Refs. \cite{warm,62526,6252602,6252603,6252604}. In this way,
utilizing Eqs.(\ref{key_02}) and (\ref{inf3}), the energy density
of the radiation field, yields
\begin{equation}
\rho_{\gamma}=\frac{\Gamma\dot{\phi}^{2}}{4H}=\frac{\Gamma(-\dot{H})}{2\kappa
H(1+R)} \left[  1-{\frac{12 H^{2}}{\kappa\rho_{c}}}\right]
^{-1/2}=C_{\gamma}\,T^{4},
\label{rh}%
\end{equation}
where the constant $C_{\gamma}=\pi^{2}\,g_{\ast}/30$ and
$g_{\ast}$ denotes the number of relativistic degrees of freedom.
Using the above expression for $\rho_{\gamma}$,  we derive  that
the temperature of the thermal bath $T$, results
\begin{equation}
T=\left[
\frac{\Gamma\,(-\dot{H})}{2\,\kappa\,\,C_{\gamma}H\,(1+R)}\right]
^{1/4}\left[  1-{\frac{12 H^{2}}{\kappa\rho_{c}}}\right]  ^{-1/8}.
\label{rh-1}%
\end{equation}

Moreover, considering, Eqs.(\ref{G}) and (\ref{rh-1}) we get that
\begin{equation}
\Gamma^{{\frac{4-m }{4}}}\,(1+R)^{\frac{m }{4}}=C_{\phi}\left[
\frac {1}{2\kappa\, C_{\gamma}}\right]  ^{\frac{m
}{4}}\,\phi^{1-m} \left[
\frac{-\dot{H}}{H}\right]  ^{\frac{m }{4}} \left[  1-{\frac{12 H^{2}}%
{\kappa\rho_{c}}}\right]  ^{-{\frac{m }{8}}}. \label{G1}%
\end{equation}
We note that Eq.(\ref{G1}) establishes  the dissipation
coefficient $\Gamma$ in the weak (or strong) dissipative regime in
terms of the scalar field (or the cosmological time).

Otherwise, the scalar potential from Eqs.(\ref{newfried}),
(\ref{inf3}) and (\ref{rh}), becomes

\begin{equation}
V={\frac{\rho_{c}}{2}}\left[  1-\sqrt{1-{\frac{12 H^{2}}{\kappa\rho_{c}}}%
}\right]  + \frac{\dot{H}}{\kappa(1+R)}\,\left(
1+\frac{3}{2}\,R\right)
\left[  1-{\frac{12 H^{2}}{\kappa\rho_{c}}}\right]  ^{-1/2}, \label{pot}%
\end{equation}
we note that this potential, could be expressed explicitly in
terms of the  field $\phi$, for the  weak (or strong) dissipative
regime.

In the following, we will study the warm-LQC model in the context
of intermediate expansion for a general form of the  dissipation
coefficient $\Gamma=C_\phi\,T^{m}/\phi^{m-1}$ for the cases $m=3$,
$m=1$. $m=0$, and $m=-1$. In our analysis, we will restrict
ourselves to the weak (or strong ) dissipation scenario.

\subsection{ The weak dissipative regime.\label{subsection1}}

In the following, we will consider that our model develops
according to the weak dissipative regime, in which $\Gamma<3H$ or
equivalently $R<1$. In this approach,  the solution for the
standard scalar field $\phi=\phi(t)$, from Eqs.(\ref{at}) and
(\ref{inf3}), becomes

\begin{equation}
\phi(t)=\phi_0+\frac{1}{B}\,\mathcal{F}[t], \label{exf}%
\end{equation}
where   the constant $B\equiv\frac{3}{2}\left(
\frac{\kappa(1-f)}{2Af}\right) ^{1/2}\left(
\frac{\kappa\,\rho_{c}}{12A^{2}f^{2}}\right) ^{f/4(1-f)}$ and the
function $\mathcal{F}[t]$ is given by the expression
\[
\mathcal{F}[t]\equiv\left(  1-\frac{12A^{2}f^{2}}{\kappa\rho_{c}t^{2(1-f)}%
}\right)  ^{3/4} \,_{2}F_{1}\left[
\frac{3}{4},\frac{4-3f}{4(1-f)},\frac
{7}{4},1-\frac{12A^{2}f^{2}}{\kappa\rho_{c}t^{2(1-f)}}\right]  ,
\]
here, $_{2}F_{1}$ is the hypergeometric function\cite{Libro} and
$\phi(t=0)=\phi_0$ is an integration constant that can be taken as
$\phi(t=0)=\phi_0=0$ (without loss of generality). Combining
Eqs.(\ref{at}) and (\ref{exf}), the Hubble parameter in terms of
the inflaton field, $\phi$, becomes $
H(\phi)=\frac{Af}{(\mathcal{F}^{-1}[B\,\phi])^{1-f}}, $ where
$\mathcal{F}^{-1}$ corresponds to the inverse  of the
hypergeometric function $\mathcal{F}$.

From Eqs.(\ref{at}),(\ref{pot})and (\ref{exf}), the scalar
potential in this scenario is given by
\begin{equation}
V(\phi)=\frac{\rho_{c}}{2}\left[  1-\sqrt{1-\frac{12A^{2}f^{2}}
{\kappa
\rho_{c}(\mathcal{F}^{-1}[B\,\phi])^{2(1-f)}}}\right]  , \label{pot11}%
\end{equation}
and now considering  Eq.(\ref{G1}) the dissipation coefficient
$\Gamma$  as a function of the inflaton field, can be written as
\begin{equation}
\Gamma(\phi)=\gamma_{0}\,\phi^{\,\frac{4(1-m)}{4-m}} \left(  \mathcal{F}%
^{-1}[B\,\phi] \right)  ^{\frac{-m}{4-m}} \left[
1-\frac{12A^{2}f^{2}}
{\kappa\rho_{c}(\mathcal{F}^{-1}[B\,\phi])^{2(1-f)}} \right]
^{\frac
{-m}{2(4-m)}}, \label{gammaph}%
\end{equation}
where the constant $\gamma_{0}=C_{\phi}^{{\frac{4}{4-m}}}\left[  \frac{(1-f)}%
{2\kappa\,C_{\gamma}}\right]  ^{\frac{m}{4-m}}$ and $m \neq4$.



On the other hand, considering  the dimensionless slow-roll
parameters, we have
$
\varepsilon\equiv-\frac{\dot{H}}{H^{2}}=\frac{1-f}{Af(\mathcal{F}^{-1}%
[B\,\phi])^{f}} , $ and
$
\eta\equiv-\frac{\ddot{H}}{H \dot{H}}=\frac{2-f}{Af(\mathcal{F}^{-1}%
[B\,\phi])^{f}}. $ So, the requirement  for inflation to occur
$\varepsilon<$1  is  satisfied when $\phi
>\frac{1}{B} \,\mathcal{F}\left[  \left(  \frac{1-f}{Af}\right)
^{1/f}\right]  $.  Also, if we consider that inflationary scenario
begins at the earliest possible stage, that occurs when
$\varepsilon=1$ (see Ref. \cite{Barrow1}), we get that the value for the scalar field at the
beginning of inflation is given by $
\phi_{1}=\frac{1}{B}\,\mathcal{F}\left[ \left(
\frac{1-f}{Af}\right) ^{1/f}\right]$.

In fact, the number of e-folds $N$ between two different values of cosmological time
$t_{1}$ and $t_{2}$, or equivalently between $\phi_{1}$ and $\phi_2$, using
Eq.(\ref{exf}) is
\begin{equation}
N=\int_{t_{1}}^{t_{2}}\,H\,dt=A\,\left[
(t_{2})^{f}-(t_{1})^{f}\right] =A\,\left[
(\mathcal{F}^{-1}[B\,\phi_{2}])^{f}-(\mathcal{F}^{-1}[B\,\phi
_{1}])^{f}\right]  . \label{N}%
\end{equation}

In the following, we will analyze  the scalar and tensor
perturbations for our  model in the  scenario $R=\Gamma/3H<1$. The
complex  treatment of scalar perturbations of the effective
Hamiltonian in LQC  can be viewed  in Refs. \cite{25,37}. This
analysis is beyond the range of our article and for simplicity we
will follow the procedure of  Refs. \cite{good,Herrera:2010yg,nn}
for warm inflation in LQC model. In this form, following Refs.
\cite{warm,good,Herrera:2010yg,nn} the
density perturbation could be written as ${\mathcal{P}_{\mathcal{R}}}%
^{1/2}=\frac{H}{\dot{\phi}}\,\delta\phi$. During  warm inflation,
a thermalized radiation component is present and the fluctuation
$\delta\phi$ is mostly  thermal rather than
quantum \cite{warm,62526,6252602,6252603,6252604}. In the weak
dissipation regime, i.e., when $R=\Gamma/3H<1$, the fluctuation of
the inflaton field, $\delta\phi$, is given by
$\delta\phi^{2}\simeq H\,T$ \cite{62526,6252602,6252603,6252604,B1}. In this way, from
Eqs. (\ref{inf3}), (\ref{rh-1}), and (\ref{G1}), the power spectrum
of the scalar perturbation
${\mathcal{P}_{\mathcal{R}}}$, results%
\begin{equation}
{\mathcal{P}_{\mathcal{R}}}={\frac{\sqrt{3\pi}}{4}}\,\left(  \frac{C_{\phi}%
}{2\kappa C_{\gamma}}\right)  ^{{\frac{1}{4-m}}}\phi^{{\frac{1-m}{4-m}}%
}H^{{\frac{11-3m}{4-m}}}(-\dot{H})^{-{\frac{(3-m)}{4-m}}}\left(
1-{\frac{12H^{2}}{\kappa\rho_{c}}}\right)  ^{{\frac{3-m}{2(4-m)}}}. \label{pd1}%
\end{equation}

Combining Eqs.(\ref{exf}) and (\ref{pd1}), we obtain  that the
power spectrum as function of the scalar field becomes
\begin{equation}
{\mathcal{P}_{\mathcal{R}}}=k_{1}\,\,\phi^{\,\frac{1-m}{4-m}}
\left( \mathcal{F}^{-1}[B\,\phi] \right)
^{\frac{2f(4-m)+m-5}{4-m}} \left[ 1-\frac{12A^{2}f^{2}}
{\kappa\rho_{c}(\mathcal{F}^{-1}[B\,\phi])^{2(1-f)}}
\right]  ^{\frac{3-m}{2(4-m)}}, \label{pd}%
\end{equation}
where the constant $k_{1}$, is given by $
k_{1}={\frac{\sqrt{3\pi}\kappa}{4}}\left(  \frac{C_{\phi}}{2\kappa C_{\gamma}%
}\right)
^{{\frac{1}{4-m}}}(A\,f)^{{\frac{8-2m}{4-m}}}\,(1-f)^{{\frac
{m-3}{4-m}}}
$%
or equivalently  ${\mathcal{P}_{\mathcal{R}}}$ in terms of the
number of e-folds $N$, can be written as
\begin{equation}
{\mathcal{P}_{\mathcal{R}}}(N)=k_{2}\,\,(\mathcal{F}[J(N)])^{\,\frac{1-m}%
{4-m}} \left(  J[N] \right)  ^{\frac{2f(4-m)+m-5}{4-m}} \left[
1-\frac
{12A^{2}f^{2}} {\kappa\rho_{c} (J[N])^{2(1-f)}} \right]  ^{\frac{3-m}{2(4-m)}%
}, \label{pd}%
\end{equation}
where $J(N)$ and $k_{2}$ are given by $ J(N)=\left[
{\frac{1+f(N-1)}{Af}} \right]  ^{\frac{1}{f}} $ and $
k_{2}=k_{1}B^{-\frac{1-m}{4-m}} $, respectively.

The scalar spectral index $n_{s}$ is defined by  $n_{s}%
-1=\frac{d\ln\,{\mathcal{P}_{R}}}{d\ln k}$. In this way, from Eqs.
(\ref{N}) and (\ref{pd}) the scalar spectral index $n_{s}$, yields%

\begin{equation}
n_{s}=1-{\frac{5-m-2f(4-m)}{Af(4-m)(\mathcal{F}^{-1}[B\,\phi])^{f}}}
+n_{2}
+n_{3}, \label{nss1}%
\end{equation}
where
$$ n_{2}={\frac{1-m }{4-m}}\sqrt{{\frac{2(1-f)}{\kappa Af}}}
{\frac {(\mathcal{F}^{-1}[B\,\phi])^{-f/2}}{\phi}}\left[
1-\frac{12A^{2}f^{2}}
{\kappa\rho_{c}(\mathcal{F}^{-1}[B\,\phi])^{2(1-f)}} \right]
^{-1/4},
$$
 and
$$
n_{3}={\frac{12Af(1-f)(3-m) }{\kappa\rho_{c} (4-m)}}(\mathcal{F}^{-1}%
[B\,\phi])^{-(2-f)} \left[  1-\frac{12A^{2}f^{2}} {\kappa\rho_{c}%
(\mathcal{F}^{-1}[B\,\phi])^{2(1-f)}} \right]  ^{-1}.
$$

This  spectral index $n_{s}$, also can be written in terms of $N$,
results
\begin{equation}
n_{s}=1-\frac{5-m-2f(4-m)}{(4-m)[1+f(N-1)]}+n_{2_N} +n_{3_N}, \label{nswr}%
\end{equation}
where
$$
 n_{2_N}=B{\frac{1-m }{4-m}}\sqrt{{\frac{2(1-f)}{\kappa Af}}}
{\frac {(J[N])^{-f/2}}{\mathcal{F}[J(N)]}}\left[
1-\frac{12A^{2}f^{2}} {\kappa \rho_{c}(J[N])^{2(1-f)}} \right]
^{-1/4},
$$
and

$$
n_{3_N}={\frac{12Af(1-f)(3-m) }{\kappa\rho_{c}
(4-m)}}(J[N])^{-(2-f)} \left[ 1-\frac{12A^{2}f^{2}}
{\kappa\rho_{c}(J[N])^{2(1-f)}} \right]  ^{-1}.
$$

On the other hand,   the generation of tensor perturbations during
the inflationary period would generate gravitational waves
\cite{Bha,Ade:2014xna}.
The spectrum of the tensor perturbations is defined by ${\mathcal{P}}%
_{g}=8\kappa(H/2\pi)^{2}$.  In order to confront this model with
observations, we need to consider the tensor-to-scalar ratio, defined
as  $r=\frac{{\mathcal{P}}_g}{{\mathcal{P}_{\mathcal{R}}}}$. In
this way, from Eq.(\ref{pd}), we found that the tensor-to-scalar
ratio $r$ is given by
\begin{equation}
r=\frac{{\mathcal{P}}_g}{{\mathcal{P}_{\mathcal{R}}}}={\frac{A^{2}
f^{2} }{2 \pi^{2} M_{p}^{2}k_{1}}} \,\phi^{\,-\frac {1-m}{4-m}}
\left( \mathcal{F}^{-1}[B\,\phi] \right) ^{-\frac{3-m}{4-m}}
\left[ 1-\frac{12A^{2}f^{2}}
{\kappa\rho_{c}(\mathcal{F}^{-1}[B\,\phi
])^{2(1-f)}} \right]  ^{-\frac{3-m}{2(4-m)}} . \label{Rk}%
\end{equation}

Now, the ratio $r$, in terms of the number of e-folds $N$,
results

\begin{equation}
r = {\frac{A^{2} f^{2} }{2 \pi^{2} M_{p}^{2}k_{2}}} (\mathcal{F}%
[J(N)])^{-\frac{1-m}{4-m}} \left(  J[N] \right)
^{-\frac{3-m}{4-m}} \left[ 1-\frac{12A^{2}f^{2}} {\kappa\rho_{c}
(J[N])^{2(1-f)}} \right]  ^{-\frac
{3-m}{2(4-m)}}. \label{Rk11}%
\end{equation}

As well, we can find a relation between the ratio $R=\Gamma/3H$
and the number of e-folds $N$. In this form, combining
Eqs.(\ref{gammaph}) and (\ref{nss1}),  we get%
\begin{equation}
R(N)={\frac{\gamma_{0}B^{-\frac{4(1-m)}{4-m}}}{3Af}}\,(\mathcal{F}%
[J(N)])^{\,\frac{4(1-m)}{4-m}} \left(  J[N]\right)
^{\frac{4-2m-f(4-m)}{4-m}} \left[  1-\frac{12A^{2}f^{2}}
{\kappa\rho_{c}(J[N])^{2(1-f)}} \right]
^{\frac{-m}{2(4-m)}}. \label{Ratio-ns}%
\end{equation}
\qquad

\begin{figure}[th]
{\vspace{0.5in}
\includegraphics[width=3.0in,angle=0,clip=true]{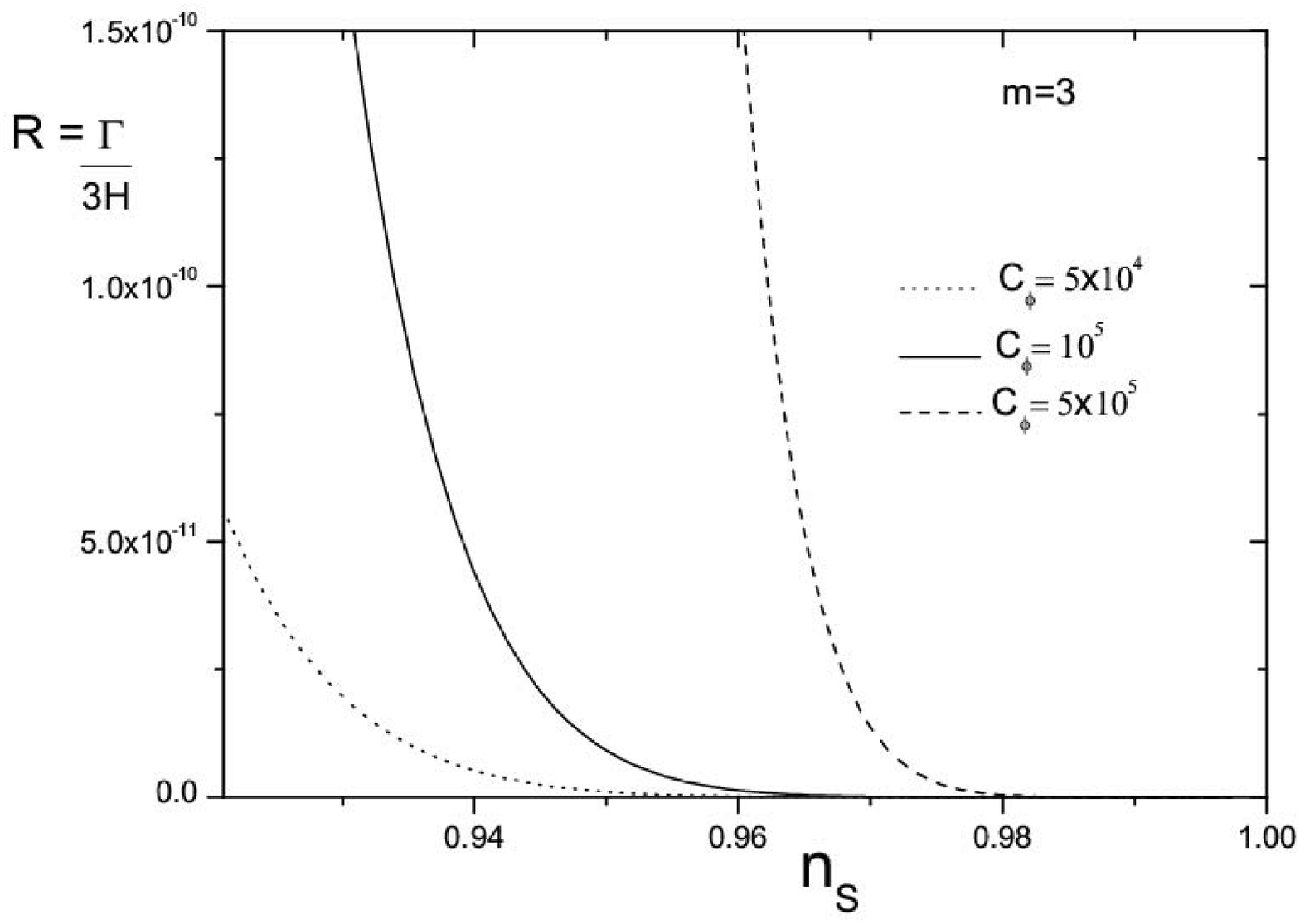}}
\includegraphics[width=3.0in,angle=0,clip=true]{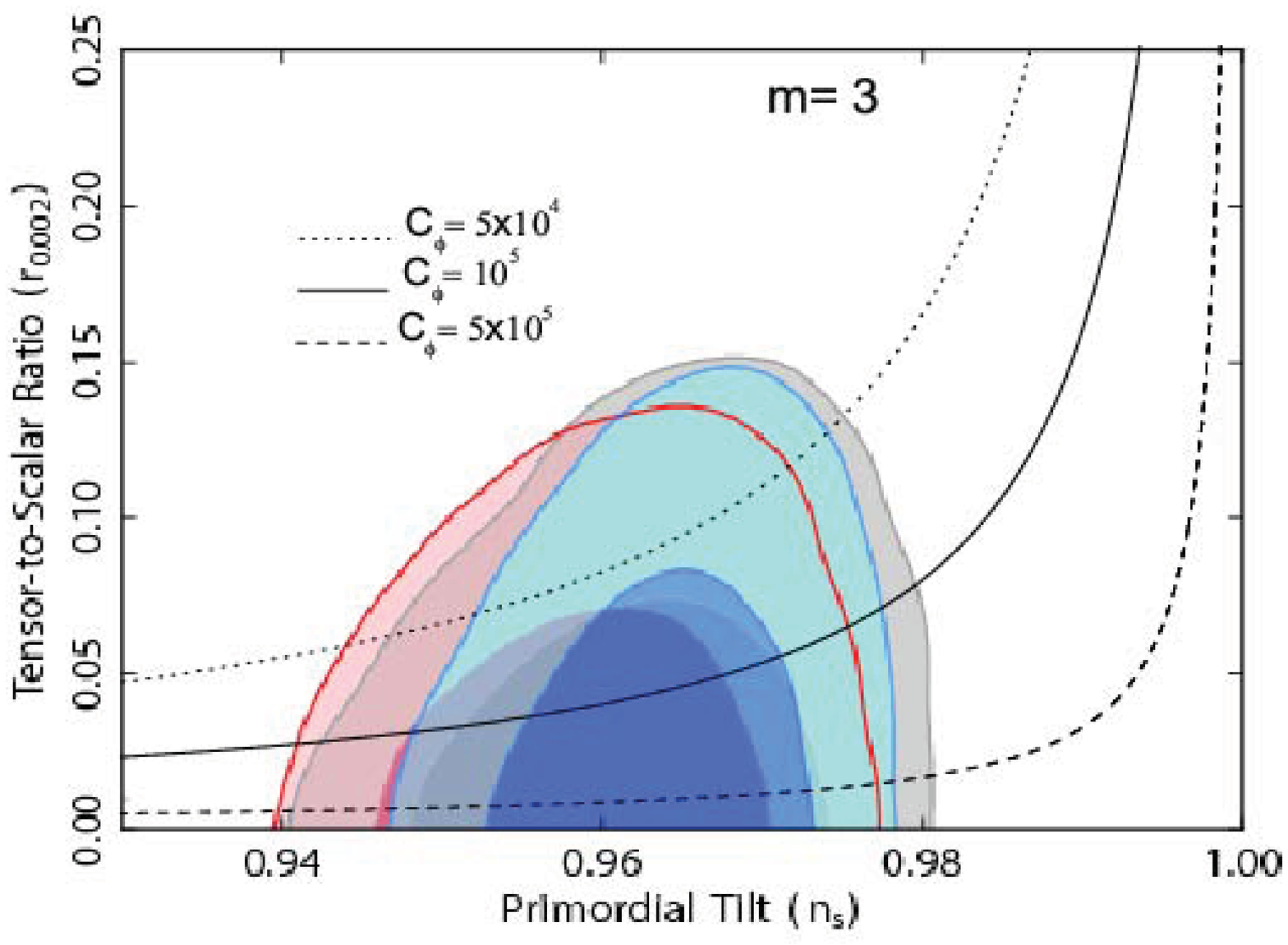}
\includegraphics[width=3.0in,angle=0,clip=true]{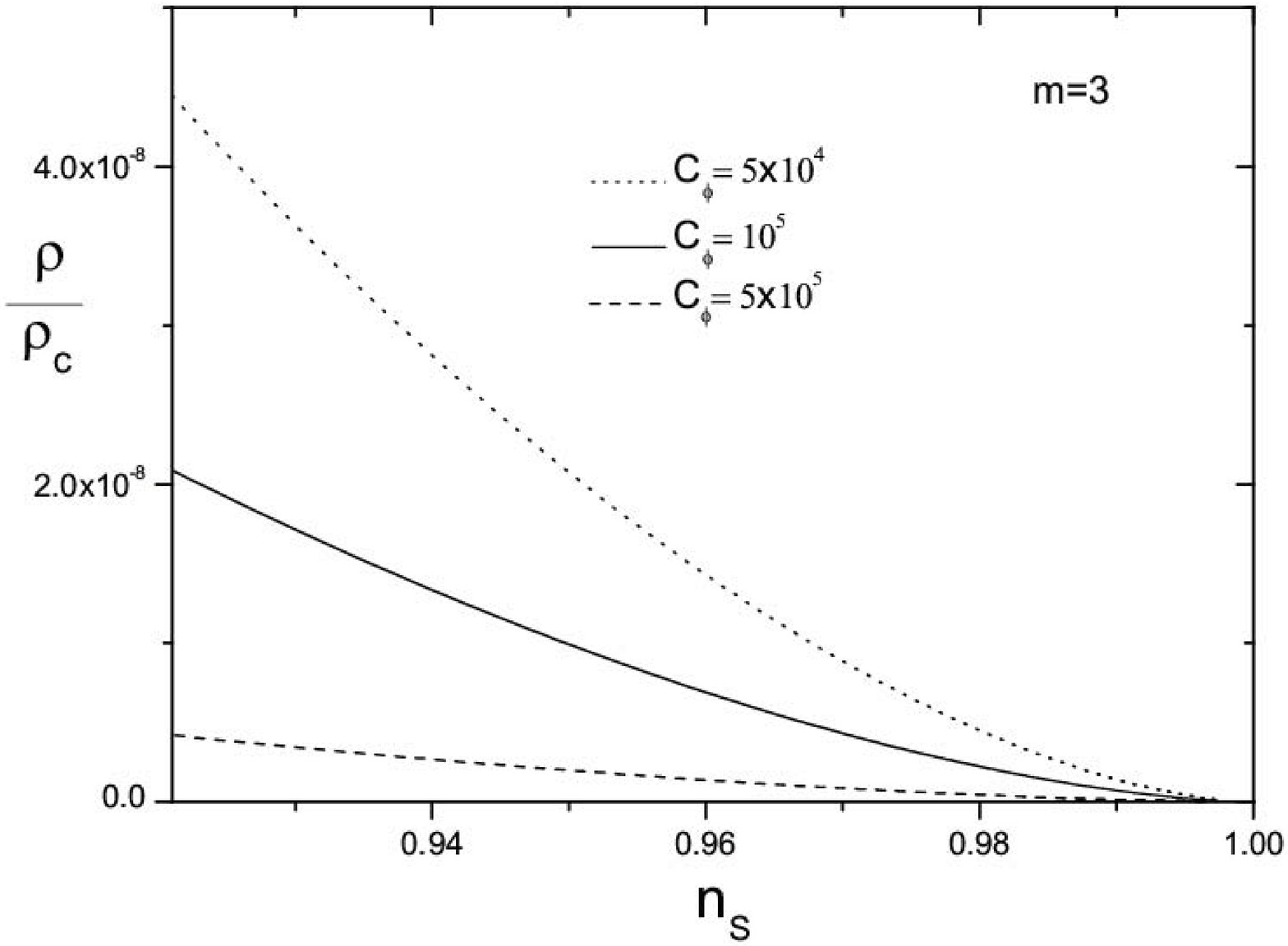}
\caption{ The evolution of the ratio $R=\Gamma/3H$ versus the
primordial tilt $n_s$ (upper left panel), the evolution of the
tensor-to-scalar ratio r versus $ n_s$ (upper right panel), and the
evolution of the ratio $\rho/\rho_c$ versus $n_s$ (lower panel) in
the warm-LQC intermediate weak dissipative regime for the case
$m=3$ in which $\Gamma\propto T^3/\phi^2$. In all panels we have
taken  three different values of the parameter $C_\phi$ and also
we have used, $\rho_c=0.82 m_p^4$,   $C_\gamma=70$, and $\kappa=1$.
In the upper right panel, we show the two-dimensional marginalized
constraints (68$\%$ and 95$\%$ CL) on inflationary parameters $r$
and $n_s$, derived from Planck data\cite{Planck}.  \label{fig1}}
\end{figure}

In Fig.\ref{fig1} we show the evolution of $R=\Gamma/3H$, the
tensor-to-scalar ratio $r$, and  the quantum geometry effects given by the ratio
$\rho/\rho_c$ on the primordial tilt $n_s$ for the special case in
which we set $m=3$  (in which $\Gamma=C_\phi\, T^3/\phi^2$), in
the warm intermediate LQC for the weak dissipative regime. In all
panels we have fixed three different values of the parameter
$C_\phi$. The upper left panel indicates the dependence  of
$R=\Gamma/3H$ on the spectral index during inflation and we also
check that the decay of the ratio $R<1$. In the upper right panel, we
exhibit the two-dimensional marginalized constraints (68$\%$ and
95$\%$ CL) from Planck data in combination with Planck + WP Planck
CMB temperature likelihood complemented  by the WMAP large-scale
polarization likelihood (grey), Planck + WP + highL (red), and
Planck + WP BAO (blue) [15]. In the lower panel we show the
development of the quantum geometry effects in LQC given by the
ratio $\rho/\rho_c$ during the inflationary scenario on the scalar
spectral index $n_s$. In order to write down values for $R$, $r$,
$\rho/\rho_c$ and $n_s$ for the case  $\Gamma\propto T^3/\phi^2$
($m =3$), we numerically manipulate  Eqs.(\ref{newfried}),
(\ref{gammaph}), (\ref{nss1}), and (\ref{Rk}), in which $C_\gamma=
70$, $\rho_c=0.82m_p^4$, and $\kappa=1$. Additionally, we
numerically solve Eqs.(\ref{pd}) and (\ref{nswr}), and we find
that $A=4.79\times 10^{-2}$ and $f=0.54$ for the case of
$C_\phi=5\times10^4$, in which $N=60$,
$\mathcal{P}_{\mathcal{R}}=2.43\times 10^{-9}$ and the scalar
spectral index $n_s=0.96$. Similarly, for the value of
$C_\phi=10^5$, corresponds to $A=3.58\times 10^{-2}$ and $f=0.55$,
and for the value of $C_\phi=5\times10^5$ corresponds to
$A=2.31\times 10^{-2}$ and $f=0.55$. From the upper left panel we
verify that the decay of the rate $R=\Gamma/3H<1$ for the
different values of the parameter $C_\phi$. From the upper right panel
we note that for $10^4<C_\phi <10^6$ the model well supported from the
Planck data for the case $m=3$, in the weak dissipative regime.
Also, from the lower panel we observe that the ratio $\rho/\rho_c$, which gives the
quantum geometry effects in LQC, becomes
$\rho/\rho_c<5\times 10^{-8}$. Here, we observe that this value
for $\rho/\rho_c$ becomes small by 2 orders of magnitude when it
is compared with the case of standard LQC, in which
$\rho/\rho_c<10^{-9}$ \cite{good}.

\begin{figure}[th]
\includegraphics[width=3.0in,angle=0,clip=true]{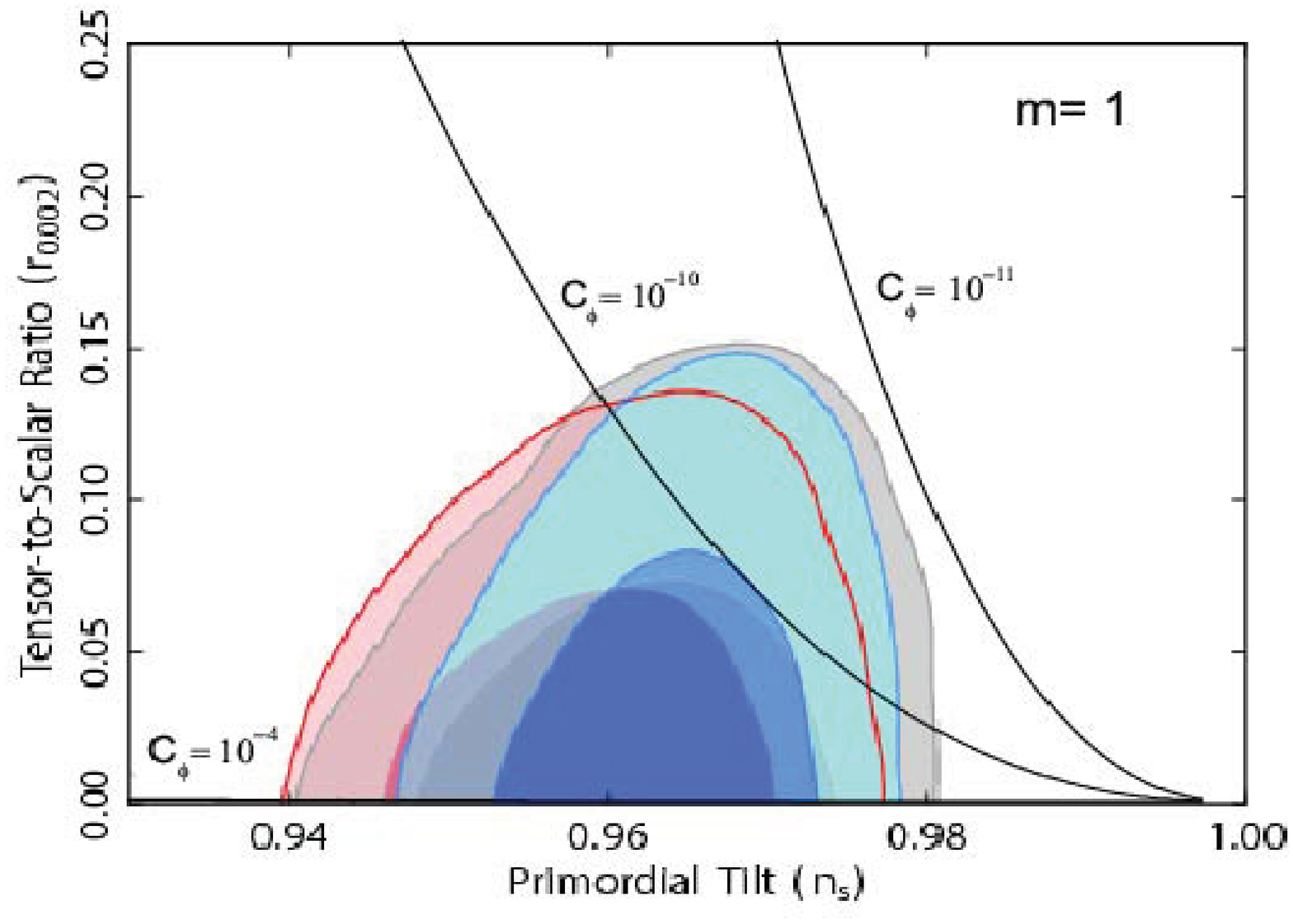}
\includegraphics[width=3.0in,angle=0,clip=true]{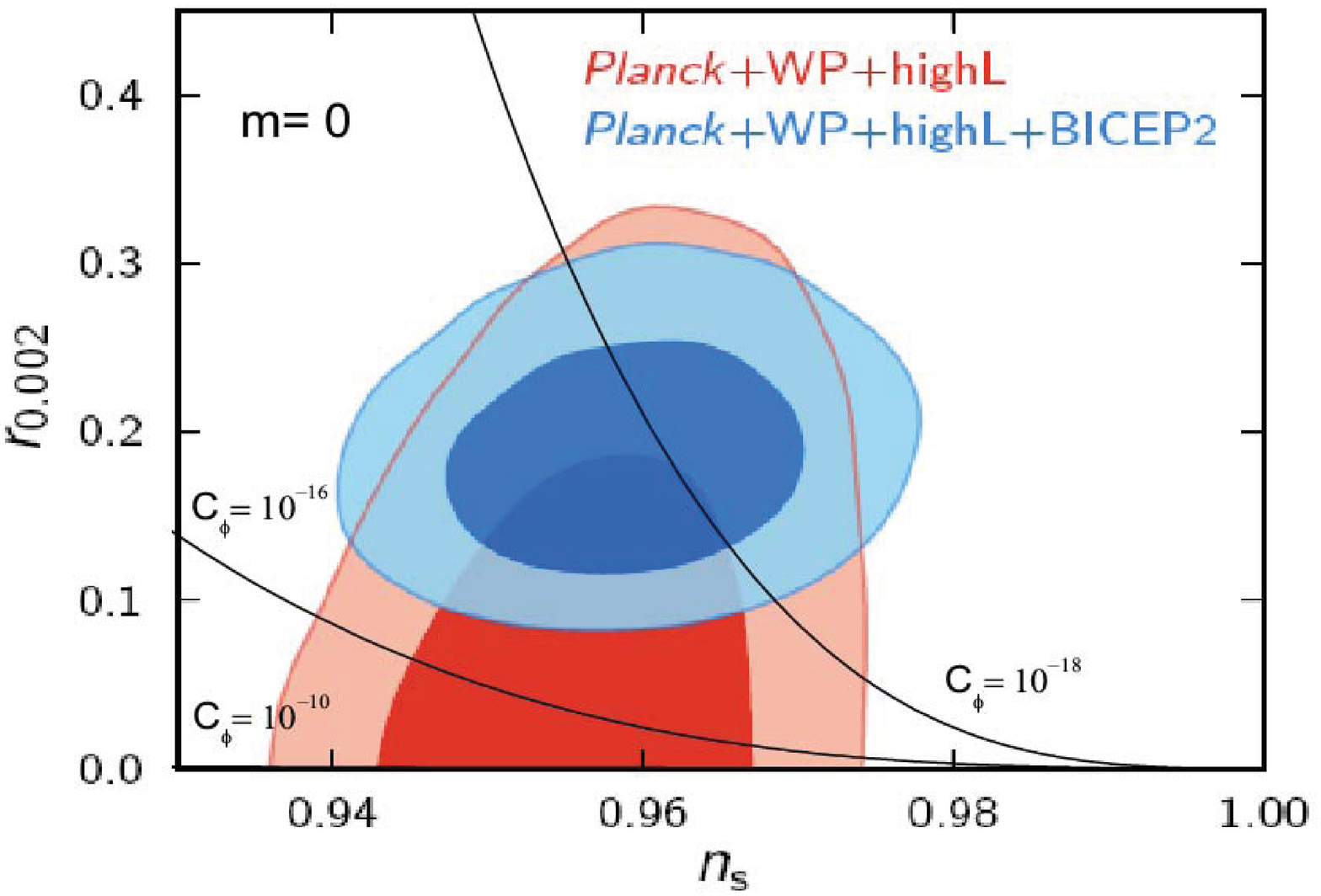}
\includegraphics[width=3.0in,angle=0,clip=true]{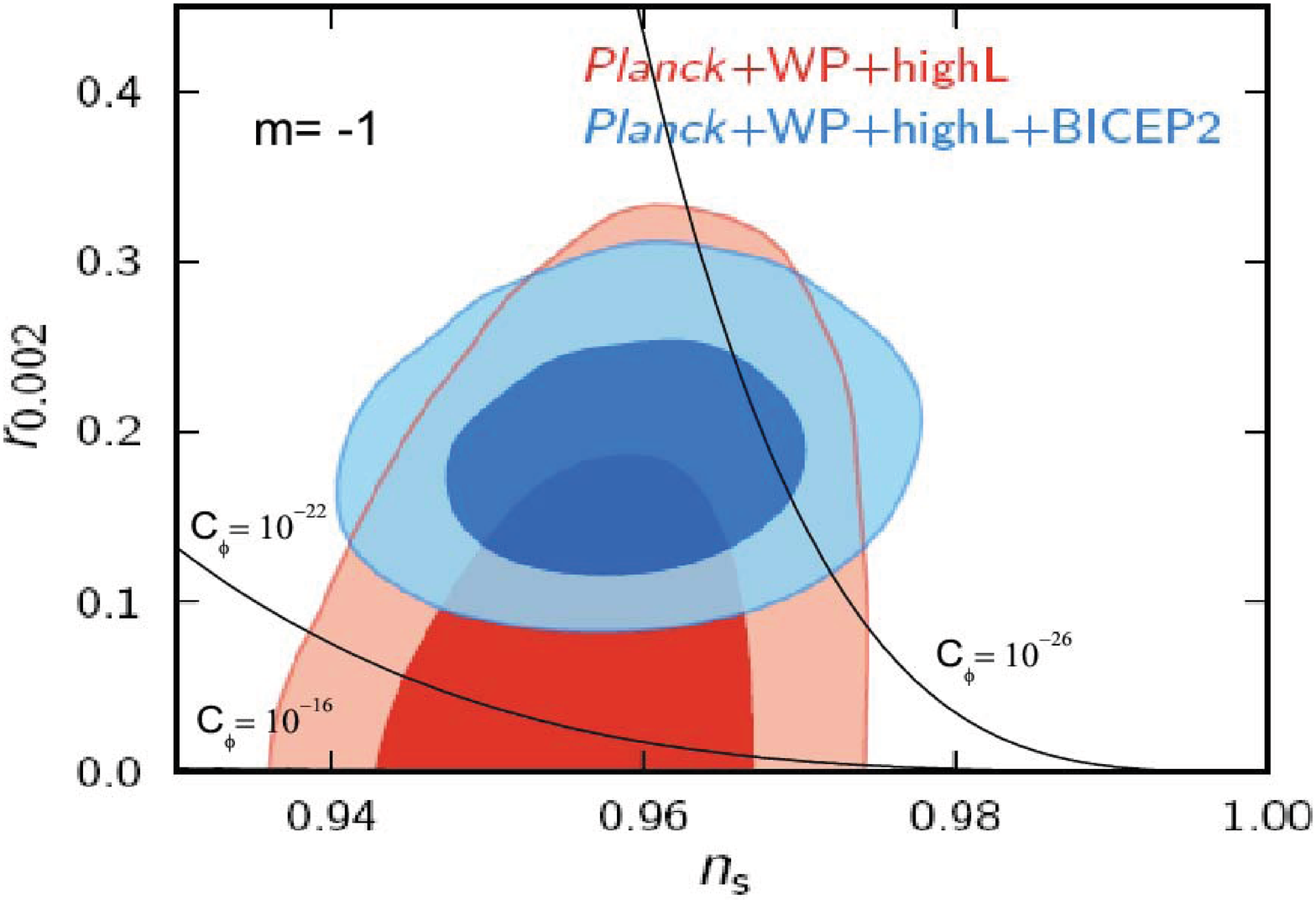}
\caption{ Evolution of the  tensor-to-scalar ratio
$r$ versus the primordial tilt $n_s$  for the cases m=1 (upper left panel),
$m=0$ (upper right panel), and $m=-1$ (lower panel) in the
warm-LQC intermediate weak dissipative regime. As before, in all
panels we have used three different values of the parameter
$C_\phi$ and also we have taken $\rho_c=0.82 m_p^4$, $C_\gamma=70$,
and $\kappa=1$. Also, in all panels, we show the
two-dimensional marginalized constraints (68$\%$ and 95$\%$ CL) on
inflationary parameters $r$ and $n_s$, derived from Planck (upper
left panel) \cite{Planck} and BICEP2 (upper right and lower panels) in
combination
 with other data sets \cite{Ade:2014xna}. \label{fig2}}
\end{figure}

In Fig.\ref{fig2} we show the evolution of the tensor-to-scalar
ratio $r$ on the spectral index $n_s$, for the cases $m=1$ , $m=0$
 and $m=-1$  in the warm LQC intermediate
 weak dissipative scenario. In all panels we use three
different values of the parameter $C_\phi$. In the upper left panel we
use $m=1$, in the upper right panel $m=0$, and in the lower panel
$m=-1$. In the upper right and lower panels, we exhibit the
two-dimensional marginalized constraints (68$\%$ and 95$\%$ CL)
from BICEP2  data in combination with Planck + WP + highL
\cite{Ade:2014xna}. We note that the BICEP2 data places stronger
limits on the tensor-to-scalar ratio $r$ versus $n_s$ compared
with the Planck data.  In order to write down values that relate
$n_s$ and $r$, as before,  we numerically solve (\ref{nss1}) and
(\ref{Rk}), where $C_\gamma= 70$, $\rho_c=0.82m_p^4$, and
$\kappa=1$. For the special case $m=1$, i.e., $\Gamma\propto T$,
we numerically solve Eqs.(\ref{pd}) and (\ref{nswr}), and we
obtain that $A=2.44$ and $f=0.28$ correspond to $C_\phi=10^{-11}$,
in which $N=60$, $\mathcal{P}_{\mathcal{R}}=2.43\times 10^{-9}$,
and the scalar spectral index $n_s=0.96$. Similarly, for the value
of $C_\phi=10^{-10}$ corresponds to $A=2.01$ and $f=0.29$, and for
$C_\phi=10^{-4}$ corresponds to $A=0.78$ and $f=0.28$.  We note
that for the value of the parameter $C_\phi> 10^{-11}$ the model
is well supported by the Planck data (upper left panel) in the warm
LQC intermediate weak regime. Here, we note that for the value of
$C_\phi=10^{-4}$ the tensor-to-scalar ratio becomes $r\sim 0$.
Also, we observe that for the case $m=1$ the value $C_\phi<
10^{-4}$ is well supported by the condition of the weak
dissipative regime, i.e., $R=\Gamma/3H<1$ (figure not shown).
Thereby, for the special case $m=1$ the constraint obtained for $C_{\phi}$ is
$10^{-11}<C_\phi<10^{-4}$.  In order to describe the quantum
geometric effect in LQC for the special case $m=1$, we
numerically find that the rate $\rho/\rho_c$ becomes $\rho/\rho_c\sim 10^{-7}$ for
 $C_\phi=10^{-11}$ evaluated at $n_s=0.96$.  For the value of the
 parameter $C_\phi=10^{-10}$ corresponds to $\rho/\rho_c\sim 10^{-8}$ and
 for value of
$C_\phi=10^{-4}$ corresponds to $\rho/\rho_c\sim 10^{-11}$ (figure
not shown).

For the  value $m=0$ in which $\Gamma\propto \phi$, as before we
numerically solve Eqs.(\ref{pd}) and (\ref{nswr}), and we find
that $A=3.12$ and $f=0.26$ correspond to $C_\phi=10^{-18}$, where
as before $N=60$, $\mathcal{P}_{\mathcal{R}}=2.43\times 10^{-9}$,
and the scalar spectral index $n_s=0.96$. Similarly, for the value
of $C_\phi=10^{-16}$ we obtain  $A=2.51$ and $f=0.25$. For the
value of  $C_\phi=10^{-10}$ we find  $A=1.21$ and $f=0.26$. As
before, we observe that for the value of the parameter
$C_\phi>10^{-19} $ the model for $m=0$ is well confirmed by the
BICEP2 data (middle panel). Here, we observe that for the value of
$C_\phi=10^{-10}$ the tensor-to-scalar ratio becomes $r\sim 0$.
Additionally,  we note that for this case of $m$ the value of the
parameter $C_\phi< 10^{-10} $ is well supported by the condition
of the weak dissipative regime, in which the rate $R=\Gamma/3H<1$
(not shown). Therefore, for the case $m=0$ we obtain for the
parameter $C_\phi$ the constraint $10^{-19}<C_\phi<10^{-16}$ from BICEP2 data.
Also, we numerically obtain that the correction term $\rho/\rho_c$
that gives the notion of the quantum geometric effects in LQC,
becomes $\rho/\rho_c\sim 10^{-8}$ for
 $C_\phi=10^{-18}$ evaluated at $n_s=0.96$.  For the value of
$C_\phi=10^{-16}$ corresponds to $\rho/\rho_c\sim 10^{-9}$ and for
$C_\phi=10^{-10}$ corresponds to $\rho/\rho_c\sim 10^{-11}$
(figure not shown).

For the  case $m=-1$ or equivalently  $\Gamma\propto \phi^2/T$, as
before we numerically solve Eqs.(\ref{pd}) and (\ref{nswr}), and
we obtain the values $A=4.2$ and $f=0.24$ for the parameter
$C_\phi=10^{-26}$. For the value of $C_\phi=10^{-22}$ corresponds
to $A=2.8$ and $f=0.23$, and for $C_\phi=10^{-16}$ corresponds to
$A=1.6$ and $f=2.4$. We find that for the value of the parameter
$C_\phi>10^{-27} $ the model for $m=-1$ is well confirmed by
BICEP2 data (lower panel). Also, we note that for this dissipation
coefficient  the value $C_\phi< 10^{-16}$ is well supported by the
condition of the weak dissipative regime, i.e., $R=\Gamma/3H<1$
(not shown). Therefore, for the special case $m=-1$, we find for
the parameter $C_\phi$ the constraint $10^{-27}<C_\phi<10^{-22}$ from BICEP2 data.
As before, we numerically get that the correction term
$\rho/\rho_c$ becomes $\rho/\rho_c<10^{-8}$, and we note that this
value is the same order of magnitude when it is compared with the
obtained by standard LQC scenario \cite{good}.

\subsection{ The strong dissipative regime.\label{subsection2}}

In this section we analyze  the strong dissipative regime
($R=\Gamma/3H>1$), and as before our model will remains in this
regime until the end of inflation. In the following, we will
consider the exact solutions for the separate  cases in which
$m=3$ and $m\neq3$. Combining Eqs.(\ref{at}), (\ref{inf3}) and
(\ref{G1}), the solution for the scalar field $\phi(t)$ in the
case $m=3$ is given by
\begin{equation}
\phi(t)-\phi_0=\exp\left[  \frac{F[t]}{C}\right]  , \label{phim3}%
\end{equation}
where the constant $C=\frac{15}{8}\left(  \frac{\kappa
C_{\phi}}{6}\right) ^{1/2}\left( \frac{3}{2\kappa
C_{\gamma}}\right)  ^{3/8}\left( \frac{1}{Af}\right)
^{5/8}(1-f)^{7/8}\left( \frac{\kappa\rho_{c}}{12A^{2}f^{2}}\right)
^{(2+5f)/16(1-f)}$. The function $F[t]$ is proportional to the
hypergeometric function and it is defined as
$F[t]\equiv \left(  1-\frac{12A^{2}f^{2}}{\kappa\rho_{c}t^{2(1-f)}%
}\right)  ^{15/16}\,_{2}F_{1}\left[  \frac{15}{16},\frac{18-11f}%
{16(1-f)},\frac{31}{16},1-\frac{12A^{2}f^{2}}{\kappa\rho_{c}t^{2(1-f)}%
}\right]   .$

For values of $m\neq3$, the new solution for the scalar field and
redefining $\varphi(t)=\frac{2}{3-m}\phi(t)^{\frac {3-m}{2}}$,
yields
\begin{equation}
\varphi(t)-\varphi_0=\frac{F_{m}[t]}{C_{m}}, \label{phimm}%
\end{equation}
where
$C_m=\left(  \frac{12+m}{8}\right)  \left(  \frac{\kappa C_{\phi}}%
{6}\right)  ^{\frac{1}{2}}\left(  \frac{3}{2\kappa
C_{\gamma}}\right) ^{\frac{m}{8}}\left(  \frac{1}{Af}\right)
^{\frac{(8-m)}{8}}(1-f)^{\frac {(m+4)}{8}}\left(
\frac{\kappa\rho_{c}}{12A^{2}f^{2}}\right)  ^{\frac{\left[
2m-4+f(8-m)\right]  }{16(1-f)}}$
 and the new function
$F_{m}[t]\equiv \left(  1-\frac{12A^{2}f^{2}}{\kappa\rho
_{c}t^{2(1-f)}}\right)  ^{(12+m)/16}\,_{2}F_{1}\left[  \frac{12+m}{16}%
,\frac{12+2m-f(8+m)}{16(1-f)},\frac{28+m}{16},1-\frac{12A^{2}f^{2}}{\kappa
\rho_{c}t^{2(1-f)}}\right]  .$  As before, without loss of
generality, we will consider $\varphi(t=0)=\varphi_{0}=0$. The
Hubble parameter as function of the inflaton field for the case
$m=3$, is given by $ H(\phi)=\frac{Af}{(F^{-1}[C\ln\phi])^{1-f}},$
and for the case in which $m\neq 3$, results $
H(\phi)=\frac{Af}{(F_{m}^{-1}[C_{m}\varphi])^{1-f}}. $ Here, in
both cases, $F^{-1}$ (or $F_m^{-1}$) represents the inverse
function of $F[t]$ (or $F_m[t]$).

The scalar potential as function of the scalar field, considering
Eqs.(\ref{pot}) and (\ref{phimm})yields
\begin{equation}
V(\phi)=\frac{\rho_{c}}{2}\left[
1-\sqrt{1-\frac{12A^{2}f^{2}}{\kappa\rho
_{c}(F^{-1}[C\ln\phi])^{2(1-f)}}}\right]  ,\,\,\,\text{ for }m=3 \label{potsr3}%
\end{equation}
and
\begin{equation}
V(\phi)=\frac{\rho_{c}}{2}\left[
1-\sqrt{1-\frac{12A^{2}f^{2}}{\kappa\rho
_{c}(F_{m}^{-1}[C_{m}\varphi])^{2(1-f)}}}\right]  ,\,\,\,\text{
for }m\neq3.
\label{potsrmm}%
\end{equation}

Analogously, as in the weak dissipative regime, the coefficient
$\Gamma=\Gamma(\phi)$ considering Eqs.(\ref{gammaph}),
(\ref{phim3}) and (\ref{phimm}) becomes
\begin{equation}
\Gamma(\phi)=\upsilon\phi^{-2}\left(  F^{-1}[C\ln\phi]\right)
^{-\frac
{3(2-f)}{4}}\left[  1-\frac{12A^{2}f^{2}}{\kappa\rho_{c}(C\ln\phi)^{2(1-f)}%
}\right]  ^{-\frac{3}{8}},\,\,\,\text{ for } m=3, \label{gammaphsr}%
\end{equation}
where $\upsilon=C_{\phi}\left(  \frac{3Af(1-f)}{2\kappa
C_{\gamma}}\right) ^{3/4}$. The dissipation coefficient for the
case $m\neq3$ is given by
\begin{equation}
\Gamma(\phi)=\upsilon_{m}\phi^{1-m}\left(
F_{m}^{-1}[C_{m}\varphi]\right)
^{-\frac{m(2-f)}{4}}\left[  1-\frac{12A^{2}f^{2}}{\kappa\rho_{c}(F_{m}%
^{-1}[C_{m}\varphi])^{2(1-f)}}\right]  ^{-\frac{m}{8}},
\label{gammaphsrmm}%
\end{equation}
where $\upsilon_{m}=C_{\phi}\left(  \frac{3Af(1-f)}{2\kappa
C_{\gamma}}\right) ^{m/4}.$

As before, for the dimensionless slow-roll parameters, we write $
\varepsilon\equiv-\frac{\dot{H}}{H^{2}}=\frac{1-f}{Af(F^{-1}[C\ln\phi])^{f}%
},\text{ for }m=3 %
$ and $
\tilde{\varepsilon}=\frac{1-f}{Af(F_{m}^{-1}[C_{m}\varphi])^{f}},\text{
for }m\neq3. $ The $\eta$ parameter becomes
$
\eta\equiv-\frac{\ddot{H}}{H\dot{H}}=\frac{2-f}{Af(F^{-1}[C\ln\phi])^{f}%
},\text{ for }m=3 $ and $
\tilde{\eta}=\frac{2-f}{Af(F_{m}^{-1}[C_{m}\varphi])^{f}},\text{
for }m\neq3. $

The inflation scenario is only satisfied when the scalar field
becomes $\phi
>\exp\left[  \frac{1}{C}\,F\left[  \left(  \frac{1-f}{Af}\right)
^{1/f}\right]  \right]  $ (for $m=3$), and $\varphi>\frac{1}{C_{m}}%
\,F_{m}\left[  \left(  \frac{1-f}{Af}\right)  ^{1/f}\right]  $
(for $m\neq3$).

The number of e-folds $N$ between two different values of the
scalar field $\phi_{1}$ and $\phi_{2}$, from Eqs.(\ref{at}),
(\ref{phim3}) and (\ref{phimm}) results in $ N =A\,\left[
F^{-1}[C\ln\phi_2])^{f}-(F^{-1}[C\ln\phi_1])^{f}\right] ,\text{
for }m=3 $ and $ N=A\,\left[
F_{m}^{-1}[C_{m}\varphi_2])^{f}-(F_{m}^{-1}[C_{m}\varphi_1
])^{f}\right]  ,\text{ for }m\neq3\text{ .} $ As in the weak
regime, we consider that the inflationary scenario begins at the
earliest possible, then the value $\phi_{1}=\exp\left[
\frac{1}{C}\,F\left[ \left( \frac{1-f}{Af}\right)  ^{1/f}\right]
\right]  $ (for the case $m=3$), and $\varphi
_{1}=\frac{1}{C_{m}}\,F_{m}\left[ \left( \frac{1-f}{Af}\right)
^{1/f}\right]  $ (for $m\neq3$).

On the other hand, as in the weak regime , the density
perturbation could be written as
${\mathcal{P}_{\mathcal{R}}}^{1/2}=\frac
{H}{\dot{\phi}}\,\delta\phi$\cite{warm}, where $\delta\phi$ in the
case of strong dissipation is given by
$(\delta\phi)^2=HT\sqrt{3R}/2\pi^2$\cite{BasteroGil:2011xd}.  In
this form, combining the Eqs.(\ref{inf3}), (\ref{rh-1}) and
(\ref{G1}) in the regime $R\gg1$, the expression for the power spectrum
of the scalar perturbation becomes
\begin{equation}
P_{\mathcal{R}}\simeq\frac{\sqrt{\pi}}{2}C_{\phi}^{\frac{3}{2}}\left(
\frac{\kappa}{6}\right)  \left(  \frac{3}{2\kappa
C_{\gamma}}\right)
^{\frac{3m+2}{8}}\phi^{\frac{3(1-m)}{2}}H^{\frac{3}{2}}\left(
-\dot
{H}\right)  ^{\frac{3m-6}{8}}\left(  1-{\frac{12H^{2}}{\kappa\rho_{c}}%
}\right)  ^{{\frac{3m-6}{8}}}. \label{Prsr}%
\end{equation}
As for the previous expressions, we need to separate the cases
$m=3$ and $m\neq3.$ Replacing Eqs.(\ref{at}), (\ref{phim3}), and (\ref{phimm}) in Eq.(\ref{Prsr}), we can express the power
spectrum  in terms of the scalar field for the two cases, and we obtain
\begin{equation}
\mathcal{P}_{\mathcal{R}}=K_{1}\phi^{-3}(F^{-1}[C\ln\phi])^{\frac{3(5f-6)}{8}%
}\left[  1-\frac{12A^{2}f^{2}}{\kappa\rho_{c}(F^{-1}[C\ln\phi])^{2(1-f)}%
}\right]  ^{-\frac{3}{16}},\text{ for }m=3, \label{Pr3}%
\end{equation}
where $K_{1}=\frac{\sqrt{\pi}}{2}C_{\phi}^{\frac{3}{2}}\left(
\frac{\kappa
}{6}\right)  \left(  \frac{3}{2\kappa C_{\gamma}}\right)  ^{\frac{11}{8}%
}(Af)^{\frac{15}{8}}(1-f)^{\frac{3}{8}}$ and \bigskip%
\begin{equation}
\mathcal{P}_{\mathcal{R}}=K_{m}\phi^{\frac{3(1-m)}{2}}(F^{-1}[C_{m}%
\varphi])^{\frac{3\left[  f(2+m)-2m\right]  }{8}}\left[  1-\frac{12A^{2}f^{2}%
}{\kappa\rho_{c}(F_{m}^{-1}[C_{m}\varphi])^{2(1-f)}}\right]
^{-\frac
{(3m-6)}{16}},\text{ for }m\neq3, \label{Prmm}%
\end{equation}
where $K_{m}=\frac{\sqrt{\pi}}{2}C_{\phi}^{\frac{3}{2}}\left(
\frac{\kappa
}{6}\right)  \left(  \frac{3}{2\kappa C_{\gamma}}\right)  ^{\frac{3m+2}{8}%
}(Af)^{\frac{3m+6}{8}}(1-f)^{\frac{3m-6}{8}}.$

By other hand, it is possible rewrite the scalar power spectrum in
terms of the number of e-folds $N$, then using Eqs.(\ref{Pr3})
and (\ref{Prmm}) we get
\begin{equation}
\mathcal{P}_{\mathcal{R}}=K_{1}\exp\left(
-\frac{3}{C}F[J[N]]\right) (J[N])^{\frac{3(5f-6)}{8}}\left[
1-\frac{12A^{2}f^{2}}{\kappa\rho
_{c}(J[N])^{2(1-f)}}\right]  ^{-\frac{3}{16}},\text{ for }m=3, \label{PRN3}%
\end{equation}
and%
\begin{equation}
\mathcal{P}_{\mathcal{R}}=K_{m}\left[  \frac{(3-m)}{2}\frac{F_{m}[J[N]]}%
{C_{m}}\right]  ^{\frac{3(1-m)}{3-m}}(J[N])^{\frac{3\left[
f(2+m)-2m\right] }{8}}\left[
1-\frac{12A^{2}f^{2}}{\kappa\rho_{c}(J[N])^{2(1-f)}}\right]
^{-\frac{(3m-6)}{16}},\text{ for }m\neq3. \label{PRNmm}%
\end{equation}

Now, the scalar spectral index  as a function of the scalar field,
considering Eqs.(\ref{at}), (\ref{phim3}), (\ref{phimm}), and (\ref{Prsr}), becomes%
\begin{equation}
n_{s}=1+{\frac{3\left(  5f-6\right)
}{8Af(F^{-1}[C\ln\phi])^{f}}}+\tilde
{n}_{2}+\tilde{n}_{3},\text{ for }m=3, \label{nsphsr}%
\end{equation}
where%
$$
\tilde{n}_{2}=-3\left(  \frac{6}{\kappa C_{\phi}}\right)  ^{\frac{1}{2}%
}\left(  \frac{3Af}{2\kappa C\phi}\right)  ^{-\frac{3}{8}}(1-f)^{\frac{1}{8}%
}(F^{-1}[C\ln\phi])^{\frac{2-3f}{8}}\left[
1-\frac{12A^{2}f^{2}}{\kappa
\rho_{c}(F^{-1}[C\ln\phi])^{2(1-f)}}\right]
^{-\frac{1}{16}}\text{ }
$$
and%
$$
\tilde{n}_{3}=-\frac{9}{2}\frac{Af(1-f)}{\kappa\rho_{c}}(F^{-1}[C\ln
\phi])^{f-2}\left[
1-\frac{12A^{2}f^{2}}{\kappa\rho_{c}(F^{-1}[C\ln
\phi])^{2(1-f)}}\right]  ^{-1}.
$$
The expression for $m\neq3$ yields%
\begin{equation}
n_{s}=1+{\frac{3\left[  f(m+2)-2m\right]  }{8Af(F_{m}^{-1}[C_{m}\varphi])^{f}%
}}+\bar{n}_{2}+\bar{n}_{3}, \label{nsphimm}%
\end{equation}
where%
$$
\bar{n}_{2}=-3\frac{(m-1)}{2}\left(  \frac{6}{\kappa
C_{\phi}}\right)
^{\frac{1}{2}}\left(  \frac{3Af}{2\kappa C\phi}\right)  ^{-\frac{m}{8}%
}(1-f)^{\frac{4-m}{8}}\frac{(F_{m}^{-1}[C_{m}\varphi])^{-\frac{\left[
m(f-2)+4\right]  }{8}}}{\phi^{\frac{(3-m)}{2}}}\times
$$
$$
\left[  1-\frac{12A^{2}f^{2}%
}{\kappa\rho_{c}(F_{m}^{-1}[C_{m}\varphi])^{2(1-f)}}\right]
^{-\frac {(4-m)}{16}},\text{ }
$$
and%
$$
\tilde{n}_{3}=-\frac{3(1-m)}{2}\frac{Af(1-f)}{\kappa\rho_{c}}(F_{m}^{-1}%
[C_{m}\varphi])^{f-2}\left[  1-\frac{12A^{2}f^{2}}{\kappa\rho_{c}(F_{m}%
^{-1}[C_{m}\varphi])^{2(1-f)}}\right]  ^{-1}.
$$
\qquad\

Analogously as in the weak regime, the scalar spectral index can be expressed in
terms of the number of e-folds $N$, obtaining

\begin{equation}
n_{s}=1+{\frac{3\left(  5f-6\right)
}{8Af(J[N])^{f}}}+\tilde{n}_{2}+\tilde
{n}_{3} ,\label{nsNtilde3}%
\end{equation}%
for the case $m=3$. Here, $\tilde{n}_{2}$  and $\tilde{n}_{3}$ are
given by
$$
\tilde{n}_{2}=-3\left(  \frac{6}{\kappa C_{\phi}}\right)  ^{\frac{1}{2}%
}\left(  \frac{3Af}{2\kappa C\phi}\right)  ^{-\frac{3}{8}}(1-f)^{\frac{1}{8}%
}(J[N])^{\frac{2-3f}{8}}\left[  1-\frac{12A^{2}f^{2}}{\kappa\rho
_{c}(J[N])^{2(1-f)}}\right]  ^{-\frac{1}{16}},
$$
and
$$
\tilde{n}_{3}=-\frac{9}{2}\frac{Af(1-f)}{\kappa\rho_{c}}(J[N])^{f-2}\left[
1-\frac{12A^{2}f^{2}}{\kappa\rho_{c}(J[N])^{2(1-f)}}\right] ^{-1}.
$$
The spectral index  for the case $m\neq3$, results
\begin{equation}
n_{s}=1+{\frac{3\left[  f(m+2)-2m\right]  }{8Af(J[N])^{f}}}+\bar{n}_{2}%
+\bar{n}_{3}, \label{nsNmm}%
\end{equation}%
where
$$
\bar{n}_{2}=-\frac{3}{8}\frac{(m-1)(12+m)}{(3-m)}\frac{(1-f)}{Af}\left(
\frac{\kappa\rho_{c}}{12(Af)^{2}}\right)  ^{\frac{(2m-4-f(m-8))}{16(1-f)}%
}\frac{(J[N])^{-\frac{\left[  m(f-2)+4\right]
}{8}}}{F_{m}[J[N]]}\left[
1-\frac{12A^{2}f^{2}}{\kappa\rho_{c}(J[N])^{2(1-f)}}\right]
^{-\frac {(4-m)}{16}}\text{ }
$$
and
$$
\tilde{n}_{3}=-\frac{3(1-m)}{2}\frac{Af(1-f)}{\kappa\rho_{c}}(J[N])^{f-2}%
\left[
1-\frac{12A^{2}f^{2}}{\kappa\rho_{c}(J[N])^{2(1-f)}}\right]
^{-1}.
$$
\begin{figure}[th]
\includegraphics[width=3.0in,angle=0,clip=true]{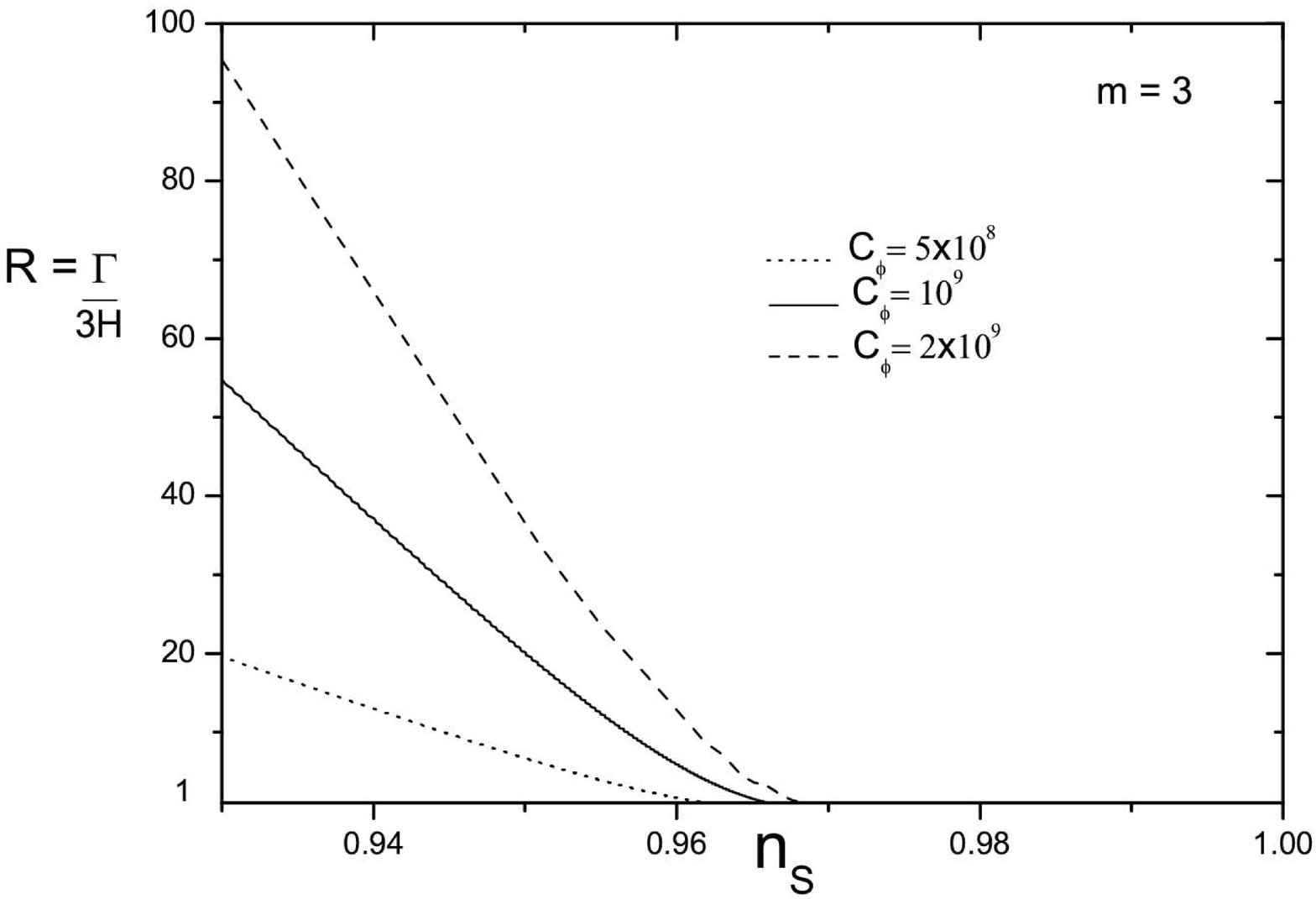}
\includegraphics[width=3.0in,angle=0,clip=true]{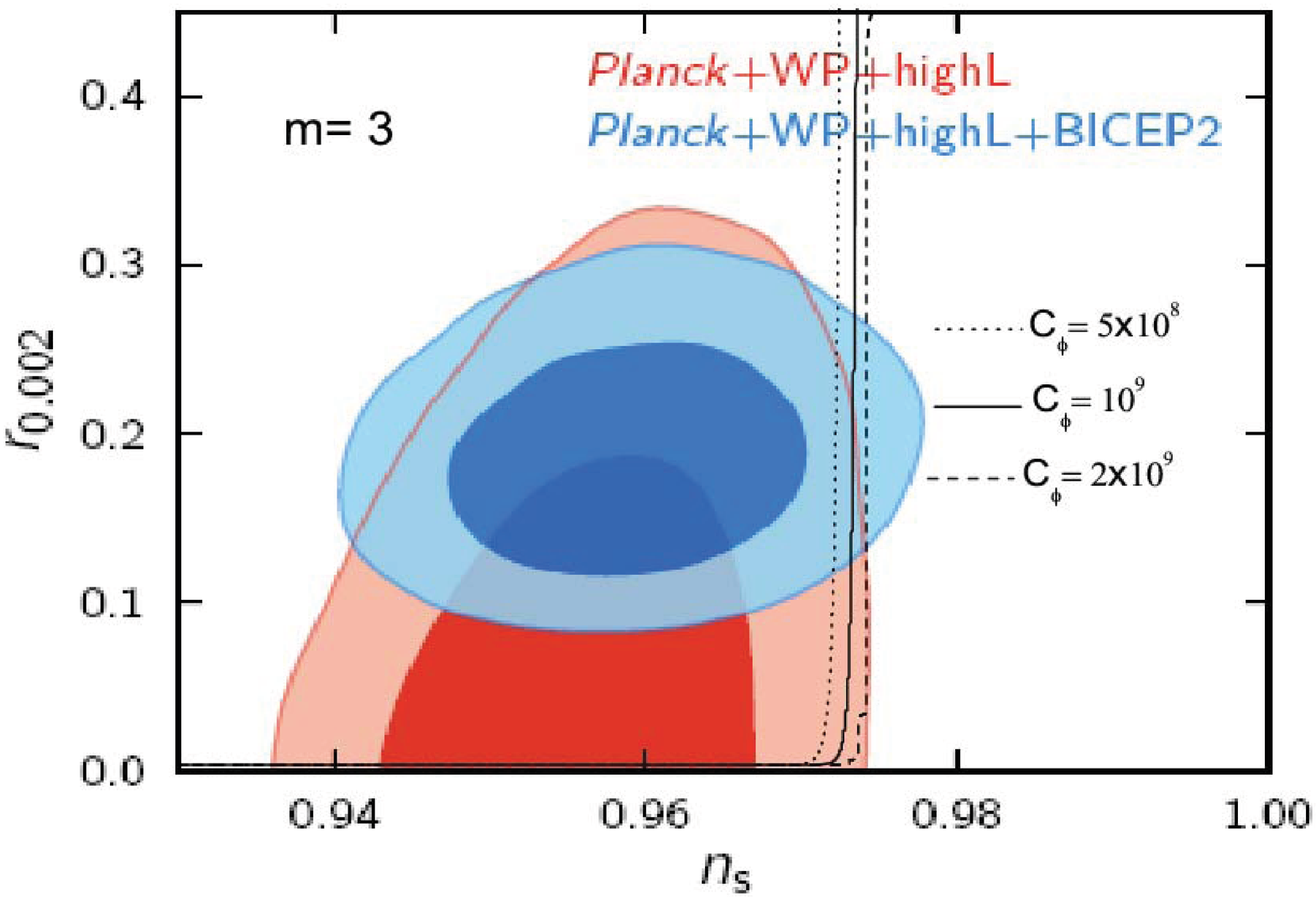}
\includegraphics[width=3.0in,angle=0,clip=true]{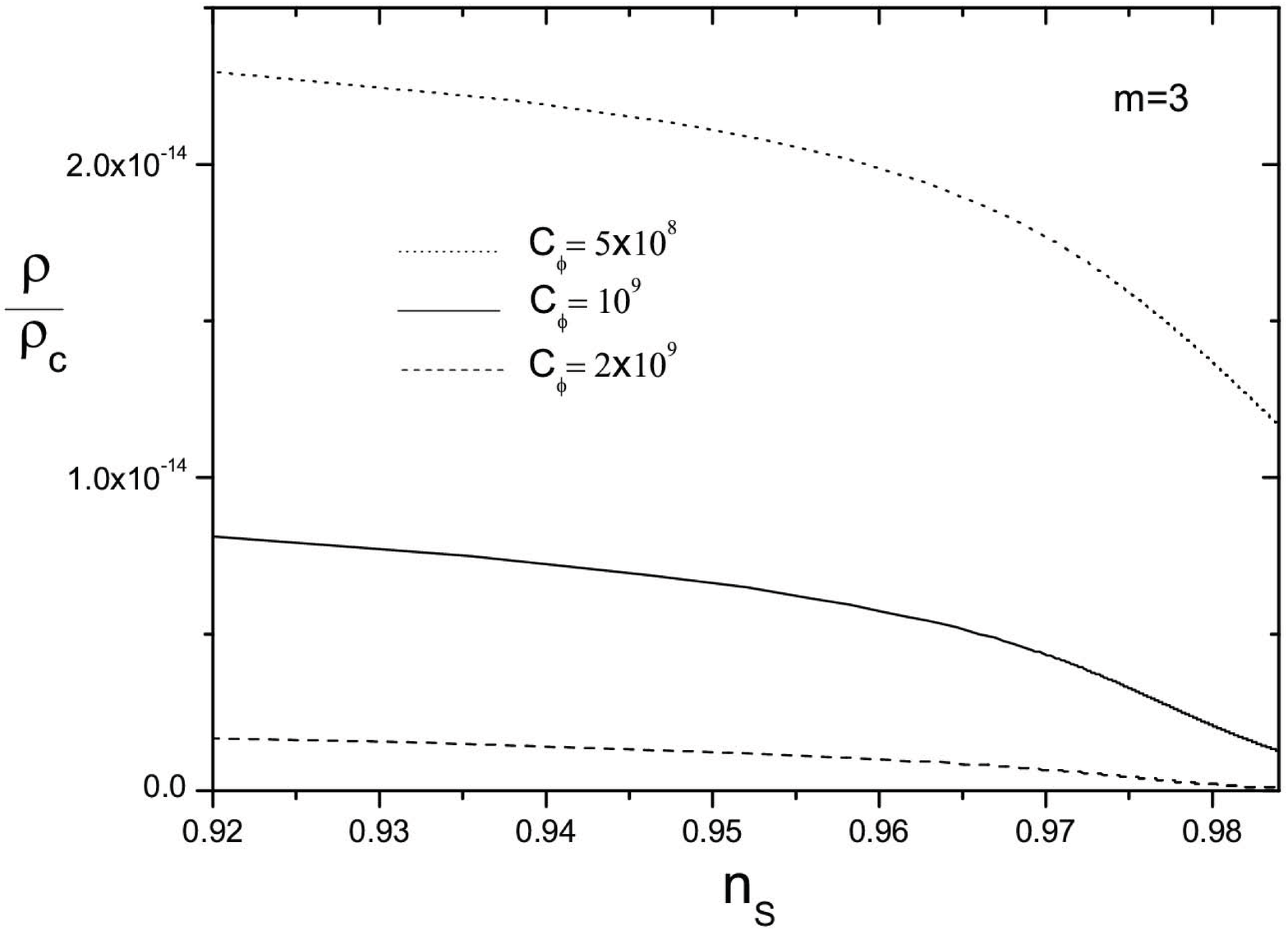}
\caption{ The evolution of the ratio $R=\Gamma/3H$ versus the
primordial tilt $n_s$ (upper left panel),  the evolution of the
tensor-to-scalar ratio r versus $ n_s$ (upper right panel), and the
evolution of the ratio $\rho/\rho_c$ versus $n_s$ (lower panel) in
the warm-LQC intermediate-strong dissipative regime for the case
$m=3$, i.e., $\Gamma\propto T^3/\phi^2$. In all panels we used
three different values of the parameter $C_\phi$ and also we have
used, $\rho_c=0.82 m_p^4$,   $C_\gamma=70$, and $\kappa=1$. In the
upper right panel, we show the two-dimensional marginalized
constraints (68$\%$ and 95$\%$ CL) on inflationary parameters $r$
and $n_s$, derived from BICEP2 data \cite{Ade:2014xna}
\label{fig5}}
\end{figure}

On the other hand, for the strong dissipative regime, the power
spectrum of the tensor perturbations is given by
${\mathcal{P}}_{g}=8\kappa(H/2\pi)^{2}$.
Using Eqs.(\ref{Pr3}) and (\ref{Prmm}) we may write the tensor-to-scalar ratio $r$ as%
\begin{equation}
r=\left(  \frac{{\mathcal{P}}_{g}}{P_{\mathcal{R}}}\right) =\frac
{A^{2}f^{2}}{2\pi^{2}M_{p}^{2}K}\phi^{3}(F^{-1}[C\ln\phi])^{\frac{f+2}{8}%
}\left[  1-\frac{12A^{2}f^{2}}{\kappa\rho_{c}(F^{-1}[C\ln\phi])^{2(1-f)}%
}\right]  ^{\frac{3}{16}},\text{ for }m=3 \label{rph3}%
\end{equation}
and%
\begin{equation}
r=\frac{A^{2}f^{2}}{2\pi^{2}M_{p}^{2}K_{m}}\phi^{\frac{3(m-1)}{2}}%
(F_{m}^{-1}[C_{m}\varphi])^{\frac{\left[  -16+f(10-3m)+6m\right]
}{8}}\left[
1-\frac{12A^{2}f^{2}}{\kappa\rho_{c}(F_{m}^{-1}[C_{m}\varphi])^{2(1-f)}%
}\right]  ^{\frac{3m-6}{16}},\text{ for }m\neq3. \label{rphmm}%
\end{equation}

Analogously, the tensor-to-scalar ratio as function of the number
of e-folds
$N$ results as%
\begin{equation}
r(N)=\frac{A^{2}f^{2}}{2\pi^{2}M_{p}^{2}K}\exp\left(  \frac{3}{C}%
F[J[N]]\right)  (J[N])^{\frac{f+2}{8}}\left[
1-\frac{12A^{2}f^{2}}{\kappa
\rho_{c}(J[N])^{2(1-f)}}\right]  ^{\frac{3}{16}},\text{ for }m=3 \label{rN3}%
\end{equation}
and%
\begin{equation}
r(N)=\frac{A^{2}f^{2}}{2\pi^{2}M_{p}^{2}K_{m}}\left[  \frac{(3-m)}{2}%
\frac{F_{m}[J[N]]}{C_{m}}\right]
^{-\frac{3(1-m)}{3-m}}(J[N])^{\frac{\left[ -16+f(10-3m)+6m\right]
}{8}}\left[  1-\frac{12A^{2}f^{2}}{\kappa\rho
_{c}(J[N])^{2(1-f)}}\right]  ^{\frac{3m-6}{16}},\text{ } \label{rNmm}%
\end{equation}
for $m\neq3.$

As in the weak regime, the analytical relation for the dissipation
ratio $R=\Gamma/3H$ between the number of e-folds is given by
\[
R(N)=\frac{\delta}{3Af}\exp\left(  -\frac{2}{C}F[J[N]]\right)
\left( J[N]\right)  ^{-\frac{(2+f)}{4}}\left[
1-\frac{12A^{2}f^{2}}{\kappa\rho _{c}(J[N])^{2(1-f)}}\right]
^{-\frac{3}{8}},\text{ for }m=3,
\]
and%
\[
R(N)=\frac{\delta}{3Af}\left[  \frac{(3-m)}{2}\frac{F_{m}[J[N]]}{C_{m}%
}\right]  ^{\frac{2(1-m)}{3-m}}\left(  J[N]\right)  ^{\frac{4(1-f)-m(2-f)}{4}%
}\left[
1-\frac{12A^{2}f^{2}}{\kappa\rho_{c}(J[N])^{2(1-f)}}\right]
^{-\frac{m}{8}},\text{ for }m\neq3.
\]

In Fig.\ref{fig5} we show the dependence of $R=\Gamma/3H$, the
tensor-to-scalar ratio $r$, and the ratio $\rho/\rho_c$ on the
primordial tilt $n_s$ for the special case in which we fix $m=3$,
in the warm LQC strong dissipative regime. In all panels we
consider three different values of the parameter $C_\phi$. In the
upper left panel we show the evolution of  $R=\Gamma/3H$ during the
inflationary epoch and we also check that the decay of the ratio
$R>1$. In the upper right panel, we exhibit the two-dimensional
marginalized constraints (68$\%$ and 95$\%$ CL) from Planck \cite{Planck} and BICEP2 in
combination which other data sets \cite{Ade:2014xna}. In the lower panel
we show the development of the quantum geometry effects in LQC
given by $\rho/\rho_c$ during the inflationary scenario. In order
to write down values for  $R$, $r$, $\rho/\rho_c$, and $n_s$ for
the value $m =3$, i.e., $\Gamma\propto T^3/\phi^2$, we manipulate numerically
the Eqs. (\ref{newfried}), (\ref{gammaphsr}), (\ref{nsphsr}), and
(\ref{rph3}) in which $C_\gamma= 70$, $\rho_c=0.82m_p^4$, and
$\kappa=1$. Additionally, we numerically solve Eqs.(\ref{PRN3})
and (\ref{rN3}) and we obtain $A=1.7\times 10^{-7}$ and $f=0.96$
for the case of $C_\phi=5\times 10^8$, in which $N=60$,
$\mathcal{P}_{\mathcal{R}}=2.43\times 10^{-9}$ and $n_s=0.96$.
Similarly, for the value of $C_\phi=10^9$, we get $A=3.6\times
10^{-7}$ and $f=0.9$, and for the value of $C_\phi=2\times 10^9$
corresponds to $A=5.2\times 10^{-7}$ and $f=0.85$. From the upper left
panel we observe that the value $C_\phi >5\times 10^{8}$ is well
confirmed by the strong regime ($R
>1$) and  this value corresponds to
an upper bound for  $C_\phi$.  From the upper right panel we observe
that for $C_\phi <2\times 10^{9}$ is well supported by the
BICEP2 data. In this form for the value $m=3$, the constraint for
the parameter $C_\phi$ becomes $5\times 10^{8}<C_\phi<2\times
10^{9}$ for the strong regime in warm intermediate model in LQC.
Also, from the lower panel we note that the quantum geometry
effects in LQC given by the ratio $\rho/\rho_c$ is
$\rho/\rho_c<3\times10^{-14}$. Additionally we note that this
inequality for $\rho/\rho_c$ becomes small by 5 orders of
magnitude when it is compared with the case of standard LQC, in
which $\rho/\rho_c<10^{-9}$ \cite{good}.

For the case $m=1$, in which $\Gamma\propto T$, we find that the
value of $C_\phi> 0.03$ is well supported by the strong
dissipative regime, i.e., $R>1$, but at the same time the
tensor-to-scalar ratio $r\sim 0$. In particular, for the value
$C_\phi=0.1$ we numerically obtain that $A=1.08$, $f=0.21$ and the
tensor-to-scalar ratio $r\simeq 5.4\times 10^{-8}$. For the other
values of $m$-parameter, we note that for the cases $m= 0$ and
$m=-1$ i.e., $\Gamma\propto \phi $ and $\Gamma\propto \phi^2/T $,
the models of the warm intermediate LQC in the strong dissipative
regime are ruled out from the Planck data and BICEP2, because
the spectral index $n_s
> 1$ and hence the models do not work.

Table I indicates the constraints on the parameter $C_\phi$ and
the quantum geometry effects in LQC given by $\rho/\rho_c$, in the
weak and strong regimes and different choices of the parameter
$m$, for a general form of $\Gamma=C_\phi\,T^{m}/\phi^{m-1}$, in
the context of warm-intermediate LCQ inflationary universe models.
\\
\\
\begin{table}
\caption{Results for the constraints on the parameter $C_\phi$ and
the quantum geometry effects in LQC given by $\rho/\rho_c$, in the
weak and strong regimes. }
\begin{tabular}
[c]{||l||l||l||l||}\hline\hline Regime &
$\Gamma=C_{\phi}\frac{T^{m}}{\phi^{m-1}}$ & \ Constraint on
$C_{\phi }$ & \,\,\,Constraint on $\frac{\rho}{\rho_{c}}$
\\\hline\hline Weak &
\begin{tabular}
[c]{c}%
$m=3$\\
$m=1$\\
$m=0$\\
$m=-1$%
\end{tabular}
&
\begin{tabular}
[c]{c}%
$5\times10^{4}<C_{\phi}<5\times10^{5}$\\
$10^{-11}<C_{\phi}<10^{-4}$\\
$10^{-19}<C_{\phi}<10^{-16}$\\
$10^{-27}<C_{\phi}<10^{-22}$%
\end{tabular}
&
\begin{tabular}
[c]{c}
$\ \ \ \ \ \ \ \ \ \ <1.47\times10^{-8}$\\
$\ \ \ \ \ \ \ \ \ \ <9.12\times10^{-8}$\\
$\ \ \ \ \ \ \ \ \ \ < 3.73\times10^{-8}$\\
$\ \ \ \ \ \ \ \ \ \ < 7.62\times10^{-8}$%
\end{tabular}
\\\hline\hline
Strong & \multicolumn{1}{||c||}{%
\begin{tabular}
[c]{c}%
$m=3$\\
$m=1$\\
$m=0$\\
$\ m=-1$%
\end{tabular}
} &
\begin{tabular}
[c]{c}%
$5\times10^{8}<C_{\phi}<2\times10^{9}$\\
$\ $The model does not work\\
$\ $The model does not work\\
$\ $The model does not work
\end{tabular}
&
\begin{tabular}
[c]{c}%
$\ \ \ \ \ \ \ \ \  < 2\times10^{-14}$\\
\ \ \ \ \ \ \ \  -\\
\ \ \ \ \ \ \ \ -\\
\ \ \ \ \ \ \ \ -
\end{tabular}
\\\hline\hline
\end{tabular}
\end{table}
\section{Conclusions \label{conclu}}

In this paper we have analyzed  the intermediate inflationary
scenario in the context of warm inflation in LQC. During the
slow-roll approximation and considering a general form of the
dissipative coefficient $\Gamma(\phi,T)=C_\phi\,T^{m/\phi^{m-1}}$,
we have found solutions of the Friedmann equations for a flat
universe  filled with a self-interacting scalar field and a
radiation field in the weak and strong dissipative regimes. In
special, we researched  the values $m=3$, $m=1$, $m=0$, and $m=-1$.
From the warm-intermediate  inflationary model in LQC, we have
found explicit relations for the corresponding scalar potential
$V(\phi)$,  spectrum of the scalar perturbations
$\mathcal{P}_{\mathcal{R}}$, scalar spectral index $n_s$, and
tensor-to-scalar ratio $r$ in the weak and strong dissipative
regimes.

In order to bring some explicit results we have considered the
constraint in the $n_s-r$ plane given by the two-dimensional
marginalized constraints (68$\%$ and 95$\%$ C.L.) derived from
Planck and BICEP2 in combinations with other data sets. Here, we noted that the BICEP2 data places
stronger limits on the tensor-to-scalar ratio $r$ versus $n_s$ compared
with the Planck data. Also, we
obtained a constraint for the value of the parameter $C_\phi$
analyzed in the weak  and strong regimes, and from these
scenarios we have found an upper bound for $C_\phi$.
Additionally, we observed that when we reduce the  parameter $m$
the value of the parameter $C_\phi$ also decreases. In particular,
for the strong dissipative regime, we found that for the cases in
which $m= 0$ and $m=-1$, i.e., for $\Gamma\propto \phi $ and
$\Gamma\propto \phi^2/T $, these models of the warm-intermediate LQC
are ruled out from Planck and BICEP2 data, since the spectral index
$n_s
> 1$, and hence the models do not work. On the other hand, for the weak
dissipative regime, the quantum geometry effects in LQC, given by
the correction term $\rho/\rho_c$ becomes similar than the
reported in the standard LQC scenario. For the strong dissipative
regime the results found indicate that the effect of the
correction term $\rho/\rho_c$ on the warm inflationary model is
marginal. Nevertheless, it cannot be rejected that future
experiments uncover it. Our results for both regimes are
summarized in Table I. Also, given that the rate $R=\Gamma/3H$
will also evolve during inflation, we may have also models which
start in the weak dissipative regime $R <1$ but end in the strong
regime, in which $R
>1$, or the other way round. In this paper, we have not studied
these dynamics. Besides, we should mention that we have not
addressed a complex treatment of the scalar perturbations of the
effective Hamiltonian in LQC, in this sense, we have considered
that the modifications to perturbation equations arise exclusively
from Hubble rate \cite{good,Herrera:2010yg,nn,g1}. We hope to
return to these points in the near future.

\begin{acknowledgments}
R.H. was supported by COMISION NACIONAL DE CIENCIAS Y TECNOLOGIA
through FONDECYT grant N$^{0}$ 1130628 and by DI-PUCV   N$^{0}$
123724.  N.V. was supported by Proyecto Beca-Doctoral CONICYT
N$^0$ 21100261.
\end{acknowledgments}



\begin{thebibliography}{99}                                                                                               %




\bibitem {R1}A. Guth , Phys. Rev. D \textbf{23}, 347 (1981).




\bibitem {R102}A.A. Starobinsky, Phys. Lett. B \textbf{91}, 99
(1980).

\bibitem {R103}A.D. Linde, Phys. Lett. B \textbf{108}, 389 (1982).

\bibitem {R104}A.D. Linde, Phys. Lett. B \textbf{129}, 177 (1983).

\bibitem {R105}A. Albrecht and P. J. Steinhardt, Phys. Rev. Lett.
\textbf{48},1220 (1982);  A. Linde, Particle Physics and
inflationary cosmology, Gordon and Breach, New York, 1990.

\bibitem {R106}K. Sato, Mon. Not. Roy. Astron. Soc. \textbf{195}, 467 (1981).

\bibitem {R2}V.F. Mukhanov and G.V. Chibisov , JETP Letters \textbf{33},
532(1981).

\bibitem {R202}S. W. Hawking,Phys. Lett. B \textbf{115}, 295
(1982).

\bibitem {R203}A. Guth and S.-Y. Pi, Phys. Rev. Lett. \textbf{49}, 1110
(1982).

\bibitem {R204}A. A. Starobinsky, Phys. Lett. B \textbf{117}, 175
(1982).

\bibitem {R205}J.M. Bardeen, P.J. Steinhardt and M.S. Turner, Phys. Rev.D
\textbf{28}, 679 (1983).


\bibitem {astro}D.~Larson \textit{et al.},
Astrophys.\ J.\ Suppl.\ \textbf{192}, 16 (2011).

\bibitem {astro2}C.~L.~Bennett \textit{et al.},
Astrophys.\ J.\ Suppl.\ \textbf{192}, 17 (2011).

\bibitem {astro202}G.~Hinshaw {\it et al.}  [WMAP Collaboration],
  Astrophys.\ J.\ Suppl.\  {\bf 208}, 19 (2013).

\bibitem{Planck}
  P.~A.~R.~Ade {\it et al.}  [Planck Collaboration],
  arXiv:1303.5082 [astro-ph.CO].

\bibitem{Ade:2014xna}
  P.~A.~R.~Ade {\it et al.}  [BICEP2 Collaboration],
  arXiv:1403.3985 [astro-ph.CO];  P. A. R. Ade
et al. [BICEP2 Collaboration], arXiv:1403.4302 [astro-ph.CO].

\bibitem {warm}A. Berera, Phys. Rev. Lett. \textbf{75}, 3218
(1995).
\bibitem {taylorberera}A. Berera, Phys. Rev. D \textbf{55}, 3346
(1997).

\bibitem {taylorberera02}J. Mimoso, A. Nunes and D. Pavon, Phys.Rev.D
\textbf{73}, 023502 (2006).

\bibitem {taylorberera03}R.~Herrera, S.~del Campo and C.~Campuzano,
JCAP \textbf{10}, 009 (2006).

\bibitem {taylorberera04}S. del Campo, R. Herrera and D. Pavon, Phys. Rev. D
\textbf{75}, 083518 (2007).

\bibitem {taylorberera05}S.~del Campo and R.~Herrera,
Phys.\ Lett.\ B \textbf{653}, 122 (2007).

\bibitem {taylorberera06}M.~A.~Cid, S.~del Campo and R.~Herrera, JCAP
\textbf{10}, 005 (2006).

\bibitem {taylorberera07}J.~C.~B.~Sanchez, M.~Bastero-Gil, A.~Berera and
K.~Dimopoulos, Phys.\ Rev.\ D \textbf{77} 123527 (2008).

\bibitem {taylorberera08}R.~Herrera,
Phys.\ Rev.\ \textbf{D81}, 123511 (2010).

\bibitem {taylorberera09}R.~Herrera, E.~San Martin,
Eur.\ Phys.\ J.\ \textbf{C71}, 1701 (2011).



\bibitem {62526}L.M.H. Hall, I.G. Moss and A. Berera, Phys.Rev.D \textbf{69},
083525 (2004).

\bibitem {6252602}I.G. Moss, Phys.Lett.B \textbf{154}, 120 (1985).

\bibitem {6252603}A.Berera and L.Z. Fang, Phys.Rev.Lett. \textbf{74} 1912
(1995).

\bibitem {6252604}A.Berera, Nucl.Phys B \textbf{585}, 666 (2000).

\bibitem {1126}A. Berera, Phys. Rev.D \textbf{54}, 2519 (1996).


\bibitem {26}I.~G.~Moss and C.~Xiong,
arXiv:hep-ph/0603266.

\bibitem {28}A.~Berera, M.~Gleiser and R.~O.~Ramos,
Phys.\ Rev.\ D \textbf{58} 123508 (1998).

\bibitem {2802}A.~Berera and R.~O.~Ramos,
Phys.\ Rev.\ D \textbf{63}, 103509 (2001).


\bibitem{Zhang:2009ge}
  Y.~Zhang,
  JCAP {\bf 0903}, 023 (2009).

\bibitem{BasteroGil:2011xd}
  M.~Bastero-Gil, A.~Berera and R.~O.~Ramos,
  JCAP {\bf 1107}, 030 (2011).


\bibitem{BasteroGil:2012cm}
  M.~Bastero-Gil, A.~Berera, R.~O.~Ramos and J.~G.~Rosa,
  JCAP {\bf 1301}, 016 (2013).


\bibitem {27}J.~C.~Bueno Sanchez, M.~Bastero-Gil, A.~Berera and
K.~Dimopoulos,
Phys.\ Rev.\ D \textbf{77}, 123527 (2008); R.~O.~Ramos and
L.~A.~da Silva,
  JCAP {\bf 1303}, 032 (2013); R.~Cerezo and J.~G.~Rosa,
  JHEP {\bf 1301}, 024 (2013).



\bibitem {Berera:2008ar}A.~Berera, I.~G.~Moss and R.~O.~Ramos,
Rept.\ Prog.\ Phys.\ \textbf{72}, 026901 (2009); M.~Bastero-Gil
and A.~Berera,
Int.\ J.\ Mod.\ Phys.\ A \textbf{24}, 2207 (2009).

\bibitem {BasteroGil:2010pb}M.~Bastero-Gil, A.~Berera and R.~O.~Ramos,
JCAP \textbf{1109}, 033 (2011); S.~del Campo and R.~Herrera,
  JCAP {\bf 0904}, 005 (2009);
R.~Herrera and M.~Olivares,
  Int.\ J.\ Mod.\ Phys.\ D {\bf 21}, 1250047 (2012);  M.~Bastero-Gil, A.~Berera, I.~G.~Moss and R.~O.~Ramos,
  arXiv:1401.1149 [astro-ph.CO].

\bibitem {PRD} J. Yokoyama and A. Linde, Phys. Rev D {\bf 60},
083509, (1999); R.~Herrera, M.~Olivares and N.~Videla,
  Phys.\ Rev.\ D {\bf 88}, 063535 (2013).


\bibitem{5} T. Thiemann ,   Lect. Notes Phys. {\bf631}, 41 (2003);  A. Ashtekar and
 J. Lewandowski ,  Class. Quant. Grav. {\bf21}, R53 (2004).

\bibitem{Ashtekar:2011ni}
   A.~Ashtekar and P.~Singh,
  Class.\ Quant.\ Grav.\  {\bf 28}, 213001 (2011)
  .

\bibitem{6} M. Bojowald ,  Living Rev. Rel. {\bf8}, 11 (2005).

\bibitem{7}M. Bojowlad ,   Phys. Rev. Lett. {\bf86}, 5227 (2001).

\bibitem{8} A. Ashtekar , M. Bojowald  and J. Lewandowski,  Adv. Theo. Math.
Phys. {\bf7}, 233 (2003).

\bibitem{9} M. Bojowald, G. Date, K. Vandersloot,  Class. Quantum Grav. {\bf21},
1253 (2004).



\bibitem{AA} A.  Ashtekar, T. Pawlowski, P. Singh  and K. Vandersloot,  Phys. Rev. D {\bf75},
024035 (2007); P.~Singh,
  Class.\ Quant.\ Grav.\  {\bf 26}, 125005 (2009)
  ; P.~Singh and F.~Vidotto,
  Phys.\ Rev.\ D {\bf 83}, 064027 (2011)
  ; P.~Singh,
  Phys.\ Rev.\ D {\bf 85}, 104011 (2012)
  ;A.~Joe and P.~Singh,
  arXiv:1407.2428 [gr-qc].

\bibitem{SinghMFE} P.~Singh,
  Phys.\ Rev.\ D {\bf 73}, 063508 (2006)
  [gr-qc/0603043].

\bibitem{As}A. Ashtekar, T. Pawlowski and P. Singh, Quantum nature of the big
bang, Phys. Rev. Lett. 96, 141301 (2006).

\bibitem{numMFE} P.~Diener, B.~Gupt and P.~Singh,
  Class.\ Quant.\ Grav.\  {\bf 31}, 105015 (2014)
  .

\bibitem{Singhinf} P.~Singh, K.~Vandersloot and G.~V.~Vereshchagin,
  Phys.\ Rev.\ D {\bf 74}, 043510 (2006)
  .


\bibitem{good} ~Zhang X. and ~Ling Y.,
JCAP {\bf 0708}, 012 (2007).

\bibitem{Ranken:2012hp}  E.~Ranken and P.~Singh,
  Phys.\ Rev.\ D {\bf 85}, 104002 (2012).


\bibitem{Gupt:2013swa} B.~Gupt and P.~Singh,
  Class.\ Quant.\ Grav.\  {\bf 30}, 145013 (2013)
 .

\bibitem{Herrera:2010yg}
  R.~Herrera,
 Phys.\ Rev.\  D {\bf 81}, 123511 (2010).


\bibitem{nn} X.~-M.~Zhang and J.~-Y.~Zhu,
  Phys.\ Rev.\ D {\bf 87}, no. 4, 043522 (2013).


\bibitem{agre1}A.~A. ~Sen,
  Phys.\ Rev.\  D {\bf 74}  043501 (2006).

\bibitem{agre2}~Xiong H.~H. and~Zhu J.~Y.,
  Phys.\ Rev.\  D {\bf 75}  084023 (2007).



\bibitem{Xiao:2011mv}
  ~Xiao K. and ~Zhu J.~Y.,
  Phys.\ Lett.\  B {\bf 699}  217 (2011).



\bibitem{int1}~Chen S., ~Wang B. and ~Jing J.,
  Phys.\ Rev.\  D {\bf 78}, 123503 (2008); ~Wu P. and ~Zhang S.~N.,
  JCAP {\bf 0806}, 007 (2008).

\bibitem{int2} J.~Mielczarek, T.~Cailleteau, J.~Grain and A.~Barrau,
  Phys.\ Rev.\ D {\bf 81}, 104049 (2010); M.~Bojowald, G.~Calcagni and S.~Tsujikawa,
  JCAP {\bf 1111}, 046 (2011);  L.~Linsefors and A.~Barrau,
  Phys.\ Rev.\ D {\bf 87}, 123509 (2013); M.~Artymowski, A.~Dapor and T.~Pawlowski,
  JCAP {\bf 1306}, 010 (2013).











\bibitem{power}F. Lucchin and S. Matarrese, Phys. Rev. D32, 1316 (1985).






\bibitem{Barrow1} J. D Barrow,
Phys. Lett. B {\bf 235}, 40 (1990); J. D Barrow and P. Saich,
Phys. Lett. B {\bf 249}, 406 (1990);A. Muslimov, Class. Quantum
Grav. {\bf 7}, 231 (1990); A. D. Rendall, Class. Quantum Grav.
{\bf 22}, 1655 (2005); J. D. Barrow and N. J. Nunes, Phys. Rev. D
\textbf{76} 043501 (2007); J. D Barrow and A. R. Liddle, Phys.
Rev. D {\bf 47}, R5219 (1993);  A. A. Starobinsky JETP Lett. {\bf
82}, 169 (2005); S.~del Campo, R.~Herrera, J.~Saavedra,
C.~Campuzano and E.~Rojas,
  Phys.\ Rev.\  D {\bf 80}, 123531 (2009);  R.~Herrera and N.~Videla,
  Eur.\ Phys.\ J.\  C {\bf 67}, 499 (2010); M.~Bastero-Gil and A.~Berera,
Int.\ J.\ Mod.\ Phys.\ A \textbf{24}, 2207 (2009);
  R.~Herrera and E.~San Martin,
  Eur.\ Phys.\ J.\ C {\bf 71}, 1701 (2011); R.~Herrera and M.~Olivares,
  Mod.\ Phys.\ Lett.\ A {\bf 27}, 1250101 (2012);  R.~Herrera, M.~Olivares and N.~Videla,
  Eur.\ Phys.\ J.\ C {\bf 73}, 2295 (2013);  R.~Herrera, M.~Olivares and N.~Videla,
  Eur.\ Phys.\ J.\ C {\bf 73}, 2475 (2013).

\bibitem{ratior}
W. H. Kinney, E. W. Kolb, A. Melchiorri and A. Riotto, Phys. Rev.
D {\bf 74}, 023502 (2006).

\bibitem{Barrow3} J. D. Barrow, A. R. Liddle and C. Pahud, Phys. Rev. D, {\bf 74}, 127305
(2006); R.~Herrera and E.~San Martin,
  Int.\ J.\ Mod.\ Phys.\ D {\bf 22}, 1350008 (2013).

\bibitem{g1} M. Bojowald, Phys. Rev. Lett. {\bf 89}, 261301
(2002); S. Tsujikawa, P. Singh and R. Maartens, Classical Quantum
Gravity {\bf 21}, 5767 (2004); E.~J.~Copeland, D.~J.~Mulryne,
N.~J.~Nunes and M.~Shaeri,
  Phys.\ Rev.\ D {\bf 77}, 023510 (2008).



\bibitem {Libro}Abramowitz, M. and Stegun, I. A. (Eds.). Handbook of
Mathematical Functions with Formulas, Graphs, and Mathematical
Tables, 9th printing. New York: Dover, 1972; Arfken, G. "The
Incomplete Gamma Function and Related Functions." Mathematical
Methods for Physicists, 3rd ed. Orlando, FL: Academic Press, 1985.

\bibitem {25} M.~Bojowald, H.~H.~Hernandez, M.~Kagan, P.~Singh and A.~Skirzewski, Phys. Rev. D 74,
123512 (2006); E.W. Ewing, Classical Quantum Gravity 29, 085005
(2012); M. Bojowald and G. M. Hossain, Phys. Rev. D 78, 063547
(2008); T. Cailleteau, J. Mielczarek, A. Barrau, and J. Grain,
Classical Quantum Gravity 29, 095010 (2012).

\bibitem {37}M.~Bojowald, G.~Calcagni and S.~Tsujikawa,
  Phys.\ Rev.\ Lett.\  {\bf 107}, 211302 (2011).




\bibitem {B1}A. Berera, Nucl. Phys. B \textbf{585}, 666 (2000).

\bibitem {Bha}K. Bhattacharya, S. Mohanty and A. Nautiyal, Phys.Rev.Lett.
\textbf{97}, 251301 (2006).





























\end{thebibliography}
\end{document}